\documentclass[draft]{agujournal2019}
\usepackage{url} 
\usepackage{lineno}
\usepackage{soul}
\usepackage{amsmath}
\usepackage{amssymb}
\usepackage{array}
\usepackage{wrapfig}
\draftfalse

\journalname{JGR: Planets}

\begin{document}

%
%


\title{Atmospheric $\text{CO}_{\text{2}}$ Ice in the Martian Polar Regions: Physical and Spectral Properties From Mars Climate Sounder Observations}

%
%




\authors{R. W. Stevens\affil{1,}\affil{2}, P. O. Hayne\affil{1,}\affil{2}, A. Kleinböhl\affil{3}, D. M. Kass\affil{4}}


\affiliation{1}{Department of Astrophysical and Planetary Sciences, University of Colorado Boulder, Boulder, CO, USA}
\affiliation{2}{Laboratory for Atmospheric and Space Physics, Boulder, CO, USA}
\affiliation{3}{Jet Propulsion Laboratory, California Institute of Technology, Pasadena, CA, USA}
\affiliation{4}{Jet Propulsion Laboratory, California Institute of Technology, Pasadena, CA, USA (retired)}




\correspondingauthor{R. W. Stevens}{robert.stevens@colorado.edu}



\begin{keypoints}
\item $\text{CO}_{\text{2}}$ ice cloud particles near the south pole of Mars have an effective radius of 46.0 $\mu$m, and an effective variance of $2.0 \times 10^{-3}$
\item There is no significant difference in observed particle size between the the northern and southern hemispheres
\item The narrow observed size distribution can be explained by Mars' tenuous, $\text{CO}_{\text{2}}$-rich atmosphere where $\text{CO}_{\text{2}}$ ice rapidly grows and precipitates
\end{keypoints}

%
%

%
%


\begin{abstract}
$\text{CO}_{\text{2}}$ ice clouds are important for polar energy balance and the carbon dioxide cycle on Mars. However, uncertainties remain regarding their physical and radiative properties, which control how polar $\text{CO}_{\text{2}}$ clouds interact with the global Martian climate. Here, we use Mars Climate Sounder (MCS) observations of atmospheric radiance to estimate these physical and radiative properties. We find that Martian $\text{CO}_{\text{2}}$ clouds are typically composed of large particles from a narrow size distribution with an effective radius of 46 $\mu$m and an effective variance of $2.0 \times 10^{-3}$ in the southern hemisphere, and an effective radius of 42 $\mu$m and an effective variance of $2.0 \times 10^{-3}$ in the north. The similarity in sizes of $\text{CO}_{\text{2}}$ ice particles in both hemispheres may be due to the fact that $\text{CO}_{\text{2}}$ clouds tend to form near the same pressure level in each hemisphere, despite the higher surface pressures in the north. We use a simplified convective cooling model to show that the small effective variance we derive may be a consequence of the fact that $\text{CO}_{\text{2}}$ is also the dominant atmospheric constituent on Mars, which allows $\text{CO}_{\text{2}}$ ice particles to reach sizes upwards of 10 $\mu$m within seconds. At the same time, the fact that the Martian atmosphere is so thin means that large particles fall rapidly to the surface, reducing the range of particle sizes that can remain in the atmosphere for any extended period of time. This study is part of ongoing work to add $\text{CO}_{\text{2}}$ ice opacity profiles to the MCS retrieval pipeline.
\end{abstract}

\section*{Plain Language Summary}
Carbon dioxide ice clouds play an important role in regulating the Martian climate. Despite their importance, there is still much we do not understand about $\text{CO}_{\text{2}}$ ice clouds, including the physical properties of the particles that form them, and how they interact with light. In this study, we use measurements of the Martian atmosphere near the poles, taken by the Mars Climate Sounder instrument, to place estimates on these ambiguous properties. We find that $\text{CO}_{\text{2}}$ clouds in both the northern and southern hemispheres are typically composed of large particles, $\sim 46$ $\mu$m, which have a very narrow distribution of sizes. The fact that that we do not observe small particles is likely due to the fact that $\text{CO}_{\text{2}}$ ice can grow very rapidly in Mars' $\text{CO}_{\text{2}}$-rich atmosphere. The lack of large particles may be a consequence of the fact that Mars’ thin atmosphere struggles to support particles once they grow large enough.

%
%

%


%
%
%
%

\section{Introduction} \label{sec:Intro}

\subsection{$\text{CO}_{\text{2}}$ Ice Clouds in the Martian Atmosphere} \label{subsec:co2atm}
The Martian poles are complex regions whose dynamics strongly influence Mars' global climate. For instance, the global $\text{CO}_{\text{2}}$ cycle, which regulates surface pressures across the entire planet, is controlled by the periodic growth and retreat of the seasonal polar caps during the winter months. Similarly, global mean surface pressures are buffered by the existence of the South Polar Residual Cap, which constitutes a permanent reservoir of $\text{CO}_{\text{2}}$ ice in solid-vapor equilibrium with the atmosphere \cite{leighton_behavior_1966}. Additionally, strong cap-edge winds, driven by the sublimation of the seasonal $\text{CO}_{\text{2}}$ ice caps, are hypothesized to play an important role in driving the annual dust storm cycle \cite{MechanismsforMarsDustStorms}.

Within the polar regions, atmospheric temperatures routinely drop below the $\text{CO}_{\text{2}}$ frost point during the winter months. When this happens in the presence of condensation nuclei, such as suspended dust or water ice particles, $\text{CO}_{\text{2}}$ ice clouds can form. These $\text{CO}_{\text{2}}$ ice clouds can strongly influence polar energy balance by backscattering infrared thermal emission, warming the surface below \cite{FORGET201381}. Additionally, snowfall driven by these polar $\text{CO}_{\text{2}}$ ice clouds is expected to contribute to the growth of the seasonal polar caps each winter alongside the direct condensation of surface frost \cite{hayne_role_2014}. Unfortunately, there is still uncertainty regarding the formation, evolution, and spatiotemporal distribution of polar $\text{CO}_{\text{2}}$ ice clouds, given that observing these atmospheric phenomena is difficult due to the lack of sunlight during the polar winter months when they form. Their full impact on these polar processes is therefore poorly understood.

Ongoing observations by the Mars Climate Sounder (MCS) onboard the Mars Reconnaissance Orbiter (MRO) spacecraft have produced a dataset representing a nearly continuous climatological record of the Martian atmosphere over nearly a full Mars decade. Currently the MCS dataset includes atmospheric profiles of temperature, pressure, and dust and water ice opacity, but lacks similar opacity profiles for $\text{CO}_{\text{2}}$ ice \cite{mccleese_mars_2007}. Previous efforts to implement $\text{CO}_{\text{2}}$ ice retrievals using the MCS dataset have been hindered by the lack of constraints on the physical and radiative properties of atmospheric $\text{CO}_{\text{2}}$ ice on Mars. In this study, we use MCS observations of atmospheric radiance to derive a best-fit particle size, particle size distribution, and infrared scattering properties for $\text{CO}_{\text{2}}$ ice clouds in the Martian polar regions, as part of ongoing work to incorporate $\text{CO}_{\text{2}}$ ice opacity profiles into the MCS data pipeline. Once incorporated, these profiles will have the potential to address many of the outstanding questions remaining in Mars polar science, including those surrounding polar energy balance and the history and evolution of the polar regions themselves.

In addition to the polar clouds already mentioned, the Martian atmosphere supports a second population of $\text{CO}_{\text{2}}$ ice clouds that form at lower latitudes $( \sim 20^{\circ}\, \text{N} - 20^{\circ} \, \text{S} )$, and at higher altitudes ($> 50$ km). These mesospheric $\text{CO}_{\text{2}}$ ice clouds are better characterized than their polar cousins, having been studied by many different authors with multiple different instruments \cite<see e.g.>[among others]{SLIPSKI2024115777, slipski_mesotides_2022, CLANCY2019246, LUGININ2024116271, AOKI2018175}. Because of the different environment in which they form, polar $\text{CO}_{\text{2}}$ ice clouds are expected to share few similarities with mesospheric $\text{CO}_{\text{2}}$ clouds, aside from composition. In this study we excluded mesospheric $\text{CO}_{\text{2}}$ clouds, as such, any future mention of ``$\text{CO}_{\text{2}}$ ice clouds" refers specifically to polar $\text{CO}_{\text{2}}$ ice clouds.

\subsection{The Importance of $\text{CO}_{\text{2}}$ Ice Clouds} \label{subsec:BG-co2}

$\text{CO}_{\text{2}}$ ice clouds were first observed as high-altitude laser echos by the Mars Orbiter Laser Altimeter (MOLA) instrument in the early 2000s \cite{neumann_two_2003}. However, early results indicated that $\text{CO}_{\text{2}}$ snowfall did not play an important role in the growth of the seasonal polar deposits, instead favoring the direct condensation of $\text{CO}_{\text{2}}$ ice as surface frost \cite{ivanov_cloud_2001}. More recent studies using MCS data, however, show that up to 20\% of the mass of the seasonal polar deposits may in fact be contributed by $\text{CO}_{\text{2}}$ snowfall \cite{hayne_role_2014}. This is significant because the granularity of an ice deposit is strongly influenced by the method of its deposition, and has a profound effect on its spectral characteristics, including its emissivity and albedo. Direct condensation of $\text{CO}_{\text{2}}$ ice as surface frost results in very coarse-grained deposits that typically have a high emissivity and low albedo, while snowfall from ice clouds above the Martian polar caps, tends to instead result in surface deposits that are much more fine-grained, with a higher albedo and lower emissivity \cite{hayne_carbon_2012, wald_modeling_1994}. Thus, a feedback occurs wherein fine-grained CO$_2$ deposits from snowfall can actually reduce total surface accumulation by decreasing the outgoing infrared emission from the surface \cite{paige1985annual, forget_low_1995, hayne_role_2014}.

Fine-grained materials are also much more sensitive to contamination by other aerosols (namely dust and water ice on Mars) due to the shorter optical path lengths between grains within the deposit. Thus, contaminants are much more likely to interact with photons in fine-grained materials, allowing a relatively small contaminant concentration to result in a disproportionately low albedo (and high emissivity) compared to coarser-grained deposits. Additionally, optically thick $\text{CO}_{\text{2}}$ snow clouds can affect the radiative environment within the poles by backscattering infrared surface emission, warming the atmosphere and the surface below \cite{hayne_carbon_2012, hinson_temperature_2004, colaprete_radiative_2000, wald_modeling_1994}. These so-called “radiatively active clouds” are beginning to be incorporated into Mars General Circulation Models (MGCMs), but their effects are highly dependent on the bulk physical properties of the $\text{CO}_{\text{2}}$ ice particles that they are composed of. A better understanding of the physical properties of $\text{CO}_{\text{2}}$ ice clouds is therefore essential to better constrain $\text{CO}_{\text{2}}$ ice cloud formation rates and snowfall amounts.

\subsection{Previous Work: CO$_2$ Ice Cloud Particle Size} \label{subsec:BG-clouds}
To date, observational constraints on the physical properties of Martian $\text{CO}_{\text{2}}$ ice come primarily from direct measurements of atmospheric radiance made by the MCS instrument. \citeA{hayne_carbon_2012} used MCS retrievals of temperature and pressure to infer the presence of atmospheric $\text{CO}_{\text{2}}$ ice from $\text{CO}_{\text{2}}$ supersaturation. Profiles that were consistent with the presence of $\text{CO}_{\text{2}}$ ice were then used to forward-model atmospheric radiances for a variety of particle sizes and cloud compositions. Radiances measured by MCS were most consistent with modeled radiances corresponding to $\text{CO}_{\text{2}}$ ice particle sizes between 10-100 $\mu$m. \citeA{hu_mars_2012} used MCS temperature profiles and MOLA-derived $\text{CO}_{\text{2}}$ ice opacity profiles to estimate the seasonal condensation area of $\text{CO}_{\text{2}}$ ice within the Martian polar regions, which was combined with estimates of the seasonal polar cap masses to derive a $\text{CO}_{\text{2}}$ ice particle size of between 8-22 $\mu$m in the north, and 4-13 $\mu$m in the south. This smaller particle size estimate is likely a consequence of the fact that MOLA is only capable of measuring the uppermost region of a cloud, where smaller ice particles are expected to be more abundant, whereas MCS is able to sample from multiple different altitude levels (and therefore particle sizes) within the cloud.

\begin{figure}[t!]
\noindent\includegraphics[width=\textwidth]{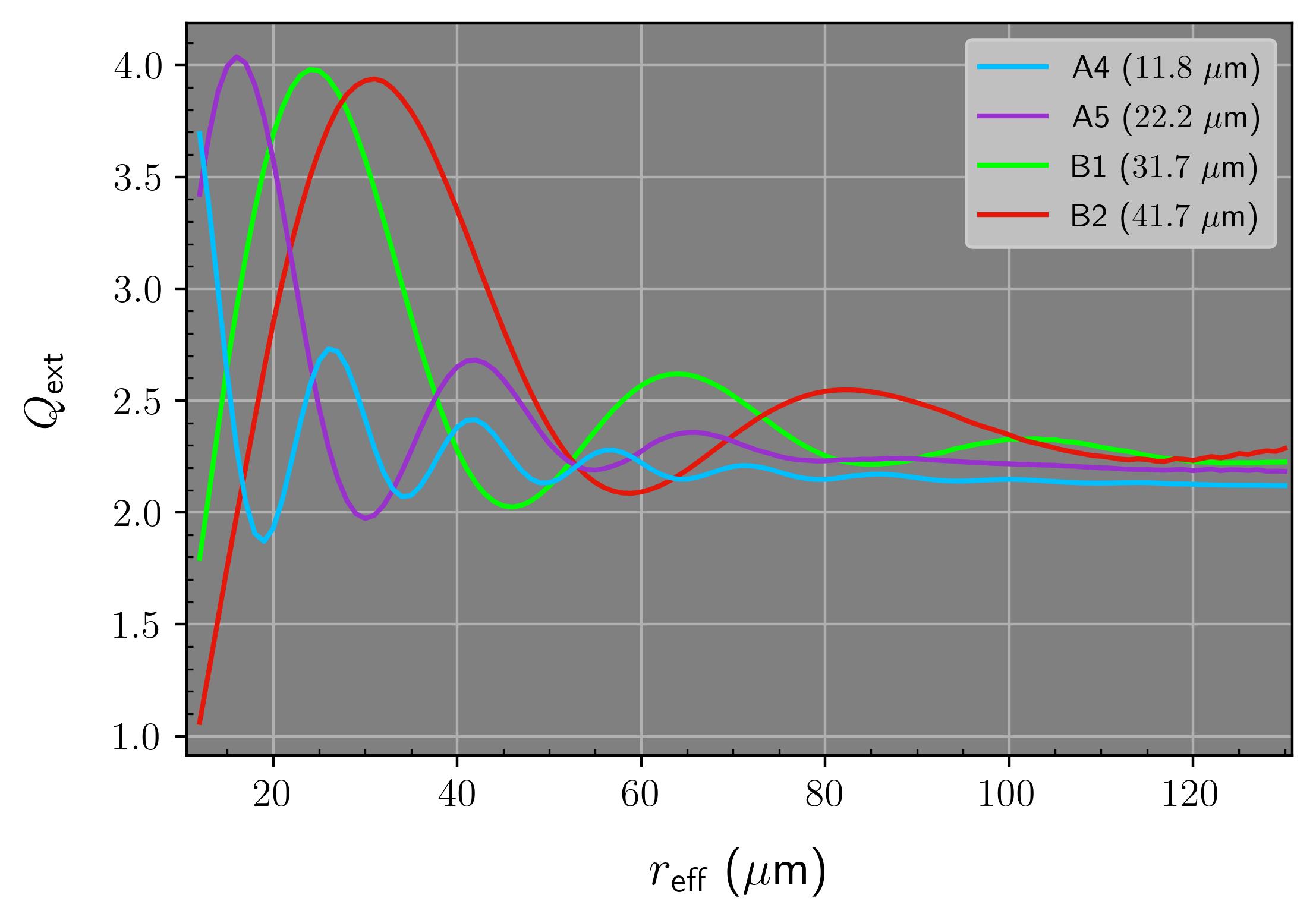}
\caption{$\text{CO}_{\text{2}}$ ice extinction efficiency as a function of particle size for the center wavelengths of each MCS aerosol channel (A4, A5, B1, B2). Values of $Q_{\text{ext}}$ can change significantly within the 10-100 $\mu$m particle size range derived by \citeA{hayne_carbon_2012}.
\label{fig:fig1}}
\end{figure}

Further estimates of $\text{CO}_{\text{2}}$ ice particle size come from modeling work using both large-scale MGCMs, and detailed microphysical models. \citeA{colaprete_carbon_2002} used a microphysical cloud model coupled to an atmospheric dynamical model to explore $\text{CO}_{\text{2}}$ snowfall, finding ice particles could grow up to 400 $\mu$m near the surface. Later, \citeA{colaprete_co2_2008} expanded their cloud model to include the effects of convection and a more realistic treatment of condensation nuclei (CN), deriving mean particle sizes of around 100 $\mu$m during polar winter, with significant seasonal and latitudinal variability. More recently, \citeA{maattanen_troposphere--mesosphere_2022} developed a detailed $\text{CO}_{\text{2}}$ cloud microphysical model that accounts for nucleation (including using water ice as CN), condensation and sublimation, and sedimentation, and incorporated this model into the Laboratoire de Météorologie Dynamique's (LMD) MGCM. Results from this study show a strong vertical and latitudinal variability in particle size, with the largest particle sizes of a few 10's of microns forming near the pole and close to the surface. The use of water ice as $\text{CO}_{\text{2}}$ cloud CN was also shown to strongly influence typical $\text{CO}_{\text{2}}$ ice particle sizes, with particles growing up to an order of magnitude larger when $\text{H}_{\text{2}}\text{O}$ ice CN are not accounted for. The relatively low $\text{CO}_{\text{2}}$ ice particle size in the more realistic model that includes water ice CN is attributed to the fact that subgrid processes such as small (sub-km)-scale convection cannot be resolved by MGCMs, and that the LMD MGCM is known to overestimate polar water ice cloud thickness, possibly contributing to an overabundance of water ice CN \cite{maattanen_troposphere--mesosphere_2022}.

While each of these studies has furthered our understanding of the microphysical properties of atmospheric $\text{CO}_{\text{2}}$ ice on Mars, the range of possible particle sizes they agree on spans at least a full order of magnitude. Given that the optical properties of $\text{CO}_{\text{2}}$ ice vary by a factor of up to $\sim$4 at infrared wavelengths within this particle size range (Figure \ref{fig:fig1}), a set of finer limits on the physical properties of polar $\text{CO}_{\text{2}}$ ice cloud particles is required before MCS data can be used to generate $\text{CO}_{\text{2}}$ ice opacity retrievals. 

In this work, we analyze a subset of MCS profiles that were generated separately with the MCS retrieval algorithm, by using the infrared scattering properties of $\text{CO}_{\text{2}}$ ice with a best-guess effective radius of 32 $\mu$m and effective variance of 0.15 \cite{hayne_carbon_2012}, instead of those of dust. Our approach is to use a forward model with particle size as a free parameter to calculate expected radiances in each of the MCS aerosol channels based on measured radiances. We use this methodology to derive a best-fit effective radius and particle size distribution. We also evaluate the sensitivity of these results to the radiative effects of water ice CN, and investigate differences in the physical properties of $\text{CO}_{\text{2}}$ ice particles between the two Martian hemispheres. Our best-fit physical properties are then used to calculate appropriate infrared scattering parameters which will ultimately be used to generate a set of $\text{CO}_{\text{2}}$ ice retrievals using a new version of the MCS retrieval algorithm.

\section{Data and Methods} \label{sec:D&M}
\subsection{Mars Climate Sounder} \label{subsubsec:MCS}

The Mars Climate Sounder (MCS) is an infrared radiometer onboard the Mars Reconnaissance Orbiter (MRO) spacecraft, designed to address key questions surrounding the Martian atmosphere and climate, including those pertaining to the global water cycle, polar energy balance, and seasonal dust storm activity \cite{mccleese_mars_2007, kass_interannual_2016}. MRO is in a sun-synchronous, nearly polar orbit, designed to provide full latitudinal coverage of the planet, and has been in orbit around Mars since March 2006. Because of its sun-synchronicity, temporal coverage is limited to local times near 3 am and 3 pm, though orbital adjustments have caused this time range to drift by no more than an hour over the course of the mission. The instrument contains nine spectral channels, eight of which (A1-A5, B1-B3) measure atmospheric radiance in wavelength bands across the mid and far infrared ($\sim$11-54 $\mu$m), that are used to retrieve vertical profiles of temperature, pressure, and aerosol opacity. The solar channel (A6) has a band pass centered at visible wavelengths (0.3-3 $\mu$m), providing information about visible reflectance and polar energy balance. However, A6 is typically not used in the atmospheric retrieval algorithm \cite{mccleese_mars_2007, kleinbohl_mars_2009}.

The MCS spectral channels each consist of an array of 21 individual detectors mounted below a set of filters that control the instrument’s spectral response. This allows MCS to perform simultaneous measurements at multiple spectral bands corresponding to specific absorption features associated with a number of gas and aerosol constiutuents found in the Martian atmosphere. As a result, each channel provides the retrieval with different information, allowing profiles of temperature, pressure, and aerosol concentration to be retrieved concurrently \cite{kleinbohl_mars_2009}.

Of particular note in MCS's design is the use of dual mechanical actuators to allow the instrument head to slew independently in both azimuth and elevation. This design feature allows the instrument to take measurements in the nadir and off-nadir directions, as well as along the planetary limb \cite{mccleese_mars_2007}. The fact that observational ray paths do not intersect with the planetary surface during limb sounding measurements is what, critically, allows MCS to measure aerosol concentrations in the Martian atmosphere; differentiating between suspended aerosols and surface deposits can be extremely difficult otherwise, as both can produce similar signals in nadir and off-nadir radiance profiles \cite{smith_separation_2000, forget_low_1995}.

In limb-sounding mode, the instrument is oriented such that each detector’s field of view intersects the Martian atmosphere at a specific limb tangent altitude, with a typical vertical resolution of $\lesssim$5 km; slightly less than a single atmospheric scale height. In this way, each array of detectors is able to measure a nearly continuous atmospheric column ranging from very near the planet’s surface to $\sim$100 km altitude. Nominally, only a subset of the full 21 detectors are used in each channel, with the retrieval algorithm excluding detectors where the measured radiance is significantly contaminated by surface thermal emission (i.e. the lowermost few detectors, including those whose lines-of-sight intersect the surface), or where the signal-to-noise is below a minimum threshold.

\subsection{MCS Retrieval Algorithm} \label{subsubsec:Ret}

A detailed discussion of the standard MCS retrieval algorithm can be found in \citeA{kleinbohl_mars_2009} and \citeA{kleinbohl_single-scattering_2011}; however, a brief overview is warranted here as well. The MCS retrieval process utilizes two primary elements: (1) a detailed 2d, single-scattering radiative transfer forward model \cite{kleinbohl2017two}; and (2) a standard atmospheric retrieval scheme based on the Chahine method of iterative relaxation \cite{chahine_general_1972} to produce atmospheric profiles that fit each set of measured radiances.

The forward model solves the radiative transfer equation to produce atmospheric radiances for a set of test temperature, pressure, and aerosol opacity profiles. Modeled radiances in each channel are controlled by the atmospheric weighting function and bulk atmospheric scattering properties along the observational line-of-sight (i.e. the single-scattering albedo and scattering phase function, which dictate the relative importance of scattering as an extinctive process, and the angular distribution of scattered light intensity, respectively). These scattering properties depend on the compositions and relative amounts of any aerosols present within the Martian atmosphere.

Calculated radiances from the forward model are compared to those measured by MCS to generate scale factors which then modify the input test profiles to provide a better fit to the measured data. This iterative relaxation process repeats multiple times until a convergence criterion is reached between the modeled and measured radiances. Because each detector channel measures a different region of the electromagnetic spectrum, individual channels are assigned to retrieve different quantities. Channels A1-A3, for instance, cover the 15 $\mu$m absorption band of gaseous $\text{CO}_{\text{2}}$, and are therefore primarily used to retrieve atmospheric temperature and pressure, while channel A4 samples the 12 $\mu$m water ice absorption feature, allowing it to retrieve water ice opacity \cite{kleinbohl_mars_2009, mccleese_mars_2007}. 
Channel A5, with a center wavelength of $\sim22 \ \mu$m, is used in the dust retrieval due to its sensitivity to atmospheric dust extinction. These final atmospheric profiles represent the ultimate data product that is uploaded to the NASA Planetary Data System.

\subsection{Dataset} \label{subsec:DataSet}

The entire MCS dataset represents a nearly continuous climatological record of the Martian atmosphere over the last nine Mars years (MY), including MY 28 and MY 34; years during which planet-encircling dust storms occurred \cite{kass_mars_2020}. The standard post-processed Level-2 data product includes atmospheric profiles of temperature, pressure, and dust and water ice opacity, as well as measured and calculated limb radiances for the specific detectors used in the retrieval. The raw retrieval output files contain pre-processed atmospheric profiles as well as calculated radiances for the full array of detectors in each channel, though radiances for detectors not used in the retrieval tend to have significantly larger uncertainties.

The atmospheric profiles used in this work are a nonstandard MCS data product where the infrared scattering properties of $\text{CO}_{\text{2}}$ ice were used during retrieval, instead of those of dust \cite[and see Section \ref{subsec:BG-clouds} of this work]{garybicas_asymmetries_2020}. Because of this, dust opacities have been replaced with $\text{CO}_{\text{2}}$ ice opacities in our dataset, even in situations where $\text{CO}_{\text{2}}$ ice is unlikely to be physically present based on local atmospheric conditions.

\begin{figure}[t!]
\noindent\includegraphics[width=\textwidth]{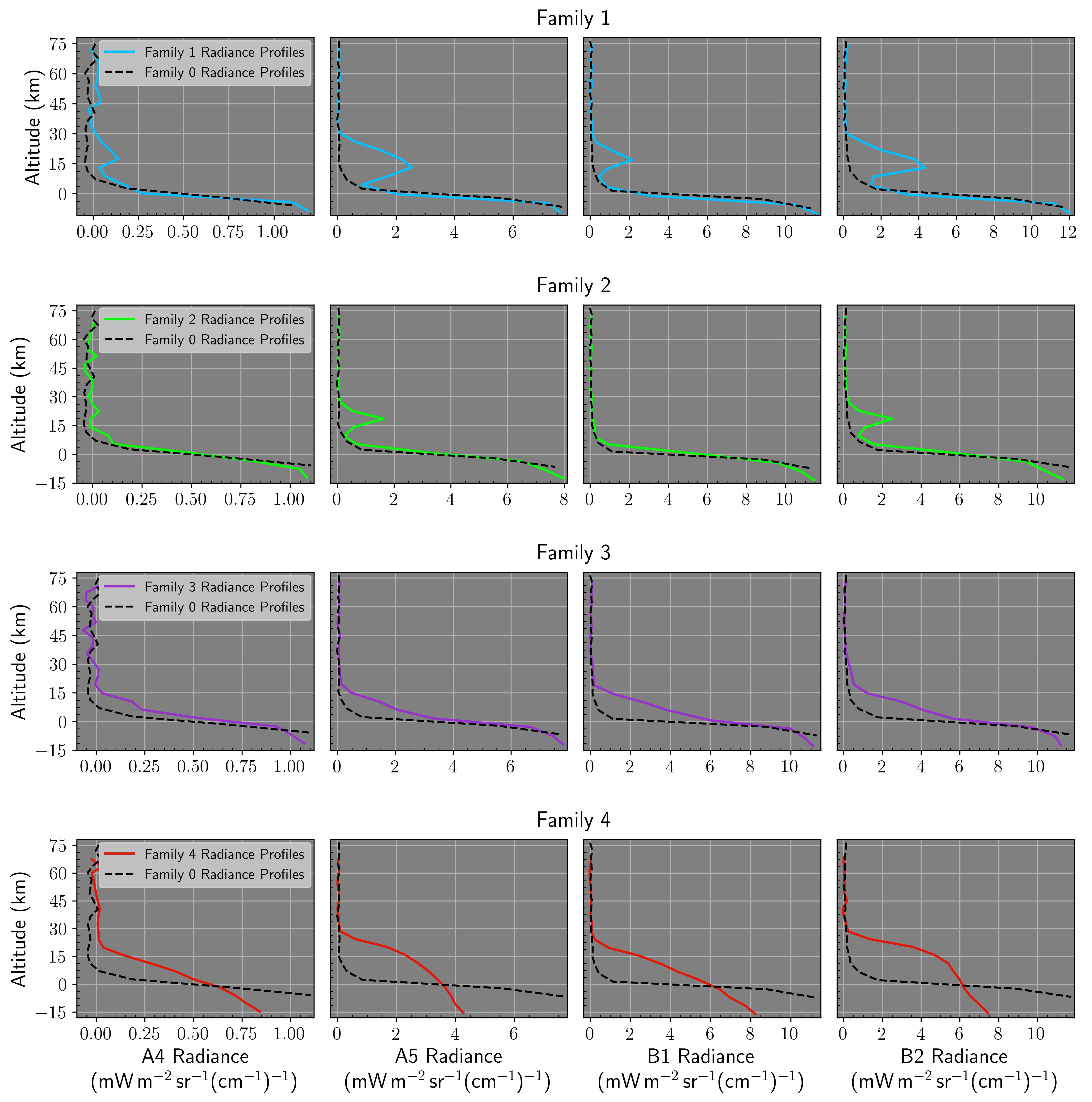}
\caption{An archetypal radiance profile from each of the south polar radiance profile ``families" identified in this work. Each row shows a different family group as it appears in each of the MCS aerosol channels. Blue lines represent the family profiles themselves, while the black dashed lines are the same in each column, and show a typical profile from ``Family 0", which corresponds to an optically thin atmosphere where very little aerosol is retrieved. An altitude of 0 km corresponds with the planet's surface. See Table \ref{tab:tab1} for a possible physical interpretation of each family.
\label{fig:fig2}}
\end{figure}

To isolate $\text{CO}_{\text{2}}$ ice clouds, we used an empirical formula from \citeA{brown_vapor_1980} to calculate the local $\text{CO}_{\text{2}}$ frost point based on retrieved atmospheric pressures. Profiles in which the temperatures did not come within 3 K of the frost point were excluded from our final analysis. The dataset was further reduced by requiring profiles be located above 65$^{\circ}$ latitude in the north, and below -65$^{\circ}$ in the south, ensuring only polar profiles were used. Because the radiance model we use (see Section \ref{subsubsec:RadMod}) requires a measured surface temperature, we additionally excluded any profiles where no surface temperature was retrieved. We also eliminated any radiances measured by detectors that were not used in the $\text{CO}_{\text{2}}$ ice retrieval itself, because detectors that are not chosen by the retrieval algorithm are either too contaminated by surface radiance or have too low signal-to-noise to be useful. We attempted to further reduce the impact of surface contamination by requiring all remaining radiances to have a minimum limb tangent altitude of 7.5 km.

The nine months of data used here were selected such that they span polar winter months in both hemispheres, near the maximum extent of the seasonal polar caps when $\text{CO}_{\text{2}}$ ice clouds regularly form. The months used in this analysis are November of 2006 ($L_s$: 128.74$^{\circ}-$143.35$^{\circ}$), July and August of 2009 ($L_s$: 295.09$^{\circ}-$327.86$^{\circ}$), December of 2014 ($L_s$: 243.83$^{\circ}-$263.39$^{\circ}$), April and October of 2015 ($L_s$: 318.09$^{\circ}-$334.68$^{\circ}$ and 48.75$^{\circ}-$62.31$^{\circ}$, respectively), and January, March, and September of 2016 ($L_s$: 89.07$^{\circ}-$100.02$^{\circ}$, 116.09$^{\circ}-$130.65$^{\circ}$, and 216.25$^{\circ}-$219.96$^{\circ}$, respectively).

Since the primary objective of this work is to determine the physical and radiative properties of $\text{CO}_{\text{2}}$ ice cloud particles, using observations where other aerosols (such as dust and water ice) are present can introduce a source of error into our results. We mitigate the impact of unretrieved dust by first focusing on observations near the south pole during the winter months (corresponding to observations from November of 2006, October of 2015, January of 2016, and March of 2016), because observed dust opacities there are typically much lower than near the north pole during northern winter \cite{garybicas_asymmetries_2020}. To reduce potential contamination by water ice, we removed any radiances where the corresponding retrieved water ice opacity at that same altitude is greater than 10$^{-5}$ km$^{-1}$, and/or the retrieved $\text{CO}_{\text{2}}$ ice opacity is less than 10$^{-3}$ km$^{-1}$.

\begin{table} [t!]
\caption{A brief description of the south polar radiance profile families identified in this work.} \label{tab:tab1}
\begin{center}
\resizebox{\textwidth}{!}{%
\begin{tabular}{  >{\centering\arraybackslash}p{3cm}  >{\centering\arraybackslash}p{9cm}  >{\centering\arraybackslash}p{9cm} }
\hline
Family Number & Family Description & Possible Physical Interpretation \\ 
\hline
0 & Low radiances in all aerosol channels (A4, A5, B1, and B2). Low $\text{CO}_{\text{2}}$ ice opacity. & Atmosphere is optically thin, no $\text{CO}_{\text{2}}$ ice clouds are present. \\
\hline
1 & Radiance peaks in each aerosol channel ($\sim 10-30$ km altitude), associated with peaks in $\text{CO}_{\text{2}}$ ice opacity profiles in same altitude range. & Detection of discrete, low altitude $\text{CO}_{\text{2}}$ ice cloud.\\
\hline
2 & Radiance peaks in A5 and B2, but no peaks in A4 and B1. Typically associated with moderate $\text{CO}_{\text{2}}$ ice opacities. & Possible detection of the edge of a small $\text{CO}_{\text{2}}$ ice cloud, potentially missed in A4 and B1 due to their positioning on the instrument.\\
\hline
3 & Smooth, shallow-sloped radiance profiles in all aerosol channels, extending from $\sim 20-30$ km to the surface. $\text{CO}_{\text{2}}$ ice opacity is moderate-to-high. & Possible detection of a $\text{CO}_{\text{2}}$ ice cloud with an extended layer of ice fog or haze below it. Possible ongoing precipitation.\\
\hline
4 & Similar in shape to Family 3 but with higher radiances and $\text{CO}_{\text{2}}$ ice opacities. Retrieval cutoff altitude where atmosphere becomes optically thick is typically above 20 km, compared to $\sim 10$ km for Family 3. & Likely detection of thick $\text{CO}_{\text{2}}$ ice cloud. Possible ongoing precipitation as $\text{CO}_{\text{2}}$ snow.\\
\hline
\end{tabular}%
}
\end{center}
\end{table}

A visual analysis of a subset of the remaining profiles from the southern polar region exposed a number of trends in the shapes of the observed radiance profiles, which we grouped into ``families" with similar characteristics. This grouping was done to ensure that profiles used in the final analysis were consistent with expectations for well-defined $\text{CO}_{\text{2}}$ ice cloud layers. Figure \ref{fig:fig2} provides an example radiance profile from each of the five families identified, and Table \ref{tab:tab1} provides a possible physical interpretation for each family. Family 1 was ultimately chosen for further analysis because it is the family for which readily apparent radiance signatures from a well-defined aerosol layer can be found in each aerosol channel. Unfortunately, after eliminating observations with significant water ice opacity, too few profiles remained to conduct a similar family grouping for the northern hemisphere. As a result, we excluded this step in our north polar particle size analysis in Section \ref{sec:NPco2}.

\subsection{Methodology} \label{subsec:Methods}


\subsubsection{Size Distributions of Atmospheric Aerosols} \label{subsubsec:SizeDist}

The modified gamma distribution has been shown to realistically model the size distributions of a number of atmospheric aerosols on Earth, including hazes, cloud particulates, and suspended ice particles \cite{deirmendjian_electromagnetic_1969}. The modified gamma distribution is, as the name implies, a variant of the well-known gamma distribution, and takes the form
\begin{equation} \label{equation:eq7}
    n \left( r \right) = a r^{\alpha} e^{-b r^{\gamma}}
\end{equation}
\noindent where $n$ is the number of particles of radius $r$ per volume element, $a$ is a normalization factor that controls the total number of particles per unit volume, $\alpha$ and $\gamma$ are two so-called ``shape parameters" that are generally determined experimentally, and $b$ is given by
\begin{equation} \label{equation:eq8}
    b = \frac{\alpha}{\gamma r_{\text{m}}^{\gamma}}
\end{equation}
\noindent where $r_{\text{m}}$ represents the mode radius of the distribution. It should be noted that both $\alpha$ and $\gamma$ must be positive, and $\alpha$ can only take integer values.

Size distributions like the modified gamma distribution are convenient because the physical ensemble of particles they represent can be described by a number of fundamental quantities obtained by taking moments of the distribution. The total number of particles per unit volume, $N$, can be calculated by taking the $0^{\text{th}}$ moment
\begin{equation} \label{equation:eq9}
    N = \int_0^{\infty} n \left( r \right) dr
\end{equation}
\noindent The effective radius, $r_{\text{eff}}$, which can be thought of as a mean radius for scattering \cite{hansen_light_1974}, is defined as the ratio of the third to the second moment of the distribution, and can be written
\begin{equation} \label{equation:eq10}
    r_{\text{eff}} = \frac{1}{G}\int_{0}^{\infty} \pi r^3 n \left( r \right) dr
\end{equation}
\noindent where
\begin{equation} \label{equation:eq11}
    G = \int_{0}^{\infty} \pi r^2 n \left(r \right) dr
\end{equation}
\noindent Finally, the spread in the distribution is controlled by the effective variance, $v_{\text{eff}}$, given by
\begin{equation} \label{equation:eq12}
    v_{\text{eff}} = \frac{1}{G r_{eff}^2} \int_{0}^{\infty} \left( r - r_{eff} \right)^2 \pi r^2 n \left(r \right) dr
\end{equation}
\noindent It has been shown that any individual modified gamma distribution can be uniquely described by $r_{\text{eff}}$ and $v_{\text{eff}}$ alone \cite{hansen_light_1974}.

\subsubsection{Modeling of Aerosol Mixtures} \label{subsubsec:AerMod}

$\text{CO}_{\text{2}}$ snowfall is capable of scavenging water ice from the Martian atmosphere by using suspended water ice crystals as condensation nuclei before falling to the planet's surface \cite{alsaeed_transport_2022}. During spring, the $\text{CO}_{\text{2}}$ ice shell surrounding each water ice core sublimates, leaving behind a layer of water ice that may eventually become incorporated into the Martian polar layered deposits. Because $\text{CO}_{\text{2}}$ ice has a transparency band around 25 $\mu$m \cite{hansen_control_1999}, it is possible that a water ice signal may be present in radiances measured by MCS, due to the presence of $\text{H}_{\text{2}}\text{O}$ ice CN, even when the retrieved water ice opacity is low. 

Furthermore, \citeA{maattanen_troposphere--mesosphere_2022} and \citeA{colaprete_formation_2003} have shown that $\text{CO}_{\text{2}}$ ice particle sizes should increase as altitude decreases. If atmospheric conditions are suitable for $\text{CO}_{\text{2}}$ condensation in multiple locations along an observational ray path, measured radiances may reflect spectral signatures from multiple size distributions with different effective radii. We thus model the radiative effects of both these physical processes by mixing Mie parameters of ices with different compositions or size distributions.

Mie parameters of aerosol mixtures are weighted by the amount, cross-sectional areas, and extinction efficiencies of each component aerosol. For an $N$-component mixture, mixed Mie parameters are given by
\begin{equation} \label{equation:eq13}
    Q_{\text{ext}_{\text{mix}}} = \frac{\displaystyle \sum_{j = 0}^N Q_{\text{ext}_j} \, n_j \, r_{\text{eff}_j}^2}{\displaystyle \sum_{j = 0}^N n_j \, r_{\text{eff}_j}^2}
\end{equation}
\begin{equation} \label{equation:eq14}
    \varpi_{0_{\text{mix}}} = \frac{\displaystyle \sum_{j = 0}^N \varpi_{0_j} \, Q_{\text{ext}_j} \, n_j \, r_{\text{eff}_j}^2}{\displaystyle \sum_{j = 0}^N Q_{\text{ext}_j} \, n_j \, r_{\text{eff}_j}^2}
\end{equation}
\begin{equation} \label{equation:eq15}
    g_{\text{mix}} = \frac{\displaystyle \sum_{j = 0}^N g_j \, \varpi_{0_j} \, Q_{\text{ext}_j} \, n_j \, r_{\text{eff}_j}^2}{\displaystyle \sum_{j = 0}^N \varpi_{0_j} \, Q_{\text{ext}_j} \, n_j \, r_{\text{eff}_j}^2}
\end{equation}
\noindent where $n_j$ is the number density of the $j^{\text{th}}$-component of the mixture, and $r_{\text{eff}}$ is the effective radius of the aerosol.

\subsubsection{Radiance Forward Model} \label{subsubsec:RadMod}

In order to constrain the particle size distribution and associated infrared scattering properties of $\text{CO}_{\text{2}}$ ice clouds, we model radiances in three MCS aerosol channels based on measured radiances in channel A5, our reference channel. Modeled radiances in each channel are then compared with measured radiances in the same channel to identify our best-fit parameters. 

Given a limb viewing geometry through a scattering atmosphere, the infrared radiances measured by MCS can be characterized as a source term times an atmospheric emissivity \cite{kleinbohl2024far}:
\begin{equation} \label{equation:eq16}
    R_{\text{m}_{\nu}} = \textbf{[} \left(1 - \varpi_{0_{\nu}} \right) B_{\nu} \left( T_{\text{eff}} \right) + \varpi_{0_{\nu}} S_{\nu} \textbf{]} \, \left( 1 - e^{-\tau_{\nu}} \right)
\end{equation}
\noindent where $R_{\text{m}}$ represents measured radiance, $\varpi_0$ is the single scattering albedo, $B$ is the Planck function at the effective limb temperature, $T_{\text{eff}}$, $S$ is the scattering source function, and $\tau$ is the optical depth. Frequency-dependent quantities are indicated by the subscript, $\nu$. The optical depth itself can be expressed as
\begin{equation} \label{equation:eqtau}
    \tau_{\nu} = \pi \, r_{\text{eff}}^2 \, Q_{\text{ext}_{\nu}} \, U
\end{equation}
\noindent where $r_{\text{eff}}$ and $Q_{\text{ext}_{\nu}}$ are the effective radius and extinction efficiency of the scattering aerosol, respectively, and $U$ is the aerosol amount in km$^{-2}$ (given by $N l$ where $N$ is the number density, and $l$ is the optical path length through the aerosol layer).

Assuming that the scattered radiance contribution is dominated by thermal emission from the surface, and that aerosol opacity is small in the nadir direction, the scattering source function can be written as
\begin{equation} \label{equation:eqscatsource}
    S_{\nu} \left( \xi, \, \theta \right) = \bar{P}_{\nu} \left( \xi, \, \theta \right) B_{\nu} \left( T_{\text{surf}} \right)
\end{equation}
\noindent where $T_{\text{surf}}$ is the surface temperature and $\bar{P}_{\nu}$ is the scattering phase function at the viewing (emission) angle, $\xi$, integrated over the full angular size of the Martian surface at a given limb tangent altitude (see \citeA{kleinbohl_single-scattering_2011} for a more detailed overview of the viewing geometry and assumptions used here). The angular size of the Martian surface is given by the surface cone half-angle, $\theta$, which can be written in terms of the tangent altitude, $z$, and planetary radius, $R_M$
\begin{equation} \label{equation:eqthetacone}
    \theta = \arcsin \left( \frac{R_M}{R_M + z} \right)
\end{equation}

Explicitly, the integrated phase function, $\bar{P}_{\nu}$, takes the form
\begin{equation} \label{equation:eqipp}
    \bar{P}_{\nu} \left( \xi, \, \theta \right) = \frac{1}{4 \pi} \int_{0}^{2 \pi} \int_{0}^{\theta} P_{\nu} \left( \psi \right) \sin \left( \theta \right) d \theta \, d \phi
\end{equation}
\noindent where $\psi$ is the scattering angle, $\phi$ is the azimuthal angle, and $P_{\nu}$ can be approximated using the single lobed Henyey-Greenstein phase function \cite{henyey1941diffuse}
\begin{equation} \label{equation:eqHGPP}
    P_{\nu} \left( \psi \right) = \frac{1 - g_{\nu}^2}{\left( 1 + g_{\nu}^2 - 2 g_{\nu} \cos \left( \psi \right) \right)^{3/2}}
\end{equation}
\noindent where $g$ is the asymmetry parameter, and the scattering angle, $\psi$, is given by
\begin{equation} \label{equation:eqpsi}
    \psi = \arccos \left( \cos \left( \xi \right) \cos \left( \theta \right) - \sin \left( \xi \right) \sin \left( \theta \right) \cos \left( \phi \right) \right)
\end{equation}

Assuming an optically thin atmosphere where $\tau_{\nu} \rightarrow 0$, and substituting in Equation \ref{equation:eqtau} for $\tau_{\nu}$ and Equation \ref{equation:eqscatsource} for $S_{\nu}$, Equation \ref{equation:eq16} can be rewritten as its first-order expansion:
\begin{equation} \label{equation:eq17}
    R_{\text{m}_{\nu}} \approx \textbf{[} \left(1 - \varpi_{0_{\nu}} \right) B_{\nu} \left( T_{\text{eff}} \right) + \varpi_{0_{\nu}} \bar{P}_{\nu} \left(\xi, \, \theta \right) B_{\nu} \left( T_{\text{surf}} \right) \textbf{]} \left( \pi \, r_{\text{eff}}^2 \, Q_{\text{ext}_{\nu}} \, U \right)
\end{equation}
\noindent Above around 1 $\mu$m in grain size, $\text{CO}_{\text{2}}$ ice is an almost perfectly scattering aerosol at wavelengths measured by the MCS aerosol channels, with a single scattering albedo very close to 1.0 \cite{hansen_spectral_1997}. This means that the scattered radiance contribution dominates in Equation \ref{equation:eq17} such that we can ignore the first term on the right. Following \citeA{kleinbohl2024far}, and assuming each MCS channel is observing the same location, Equation \ref{equation:eq17} can be used to model expected atmospheric radiances in any spectral channel given measured radiances in a different channel
\begin{equation} \label{equation:eq18}
    \frac{R_{\text{ch2}_{\text{e}}}}{R_{\text{ch1}_{\text{m}}}} = \frac{\varpi_{0_{\nu_{\text{ch2}}}} \ \bar{P}_{\nu_{\text{ch2}}} \left( \xi, \, \theta \right) \ B_{\nu_{\text{ch2}}} \left( T_{\text{surf}} \right) \, \left( \pi \, r_{\text{eff}}^2 \, Q_{\text{ext}_{\nu_{\text{ch2}}}} \, U \right)}{\varpi_{0_{\nu_{\text{ch1}}}} \ \bar{P}_{\nu_{\text{ch1}}} \left( \xi, \, \theta \right) \ B_{\nu_{\text{ch1}}} \left( T_{\text{surf}} \right) \, \left( \pi \, r_{\text{eff}}^2 \, Q_{\text{ext}_{\nu_{\text{ch1}}}} \, U \right)}
\end{equation}
\noindent Here, “ch1” and “ch2” refer to two generic spectral channels with different center frequencies, while the subscripts, ``e" and ``m", indicate expected and measured quantities, respectively. Solving for the expected radiance in channel 2, $R_{\text{ch2}_{\text{e}}}$, we get
\begin{equation} \label{equation:eq19}
    R_{\text{ch2}_{\text{e}}} = \frac{\varpi_{0_{\nu_{\text{ch2}}}} \ \bar{P}_{\nu_{\text{ch2}}} \left( \xi, \, \theta \right) \ B_{\nu_{\text{ch2}}} \left( T_{\text{surf}} \right) \ Q_{\text{ext}_{\nu_{\text{ch2}}}}}{\varpi_{0_{\nu_{\text{ch1}}}} \ \bar{P}_{\nu_{\text{ch1}}} \left( \xi, \, \theta \right) \ B_{\nu_{\text{ch1}}} \left( T_{\text{surf}} \right) \ Q_{\text{ext}_{\nu_{\text{ch1}}}}} \, R_{\text{ch1}_{\text{m}}}
\end{equation}

Using Equation \ref{equation:eq19}, we can calculate expected radiances in each of the MCS aerosol channels (A4, B1, B2) using infrared scattering properties ($Q_{\text{ext}}$, $\varpi_0$, $g$) associated with a number of different particle size distributions. The best-fit scattering properties, and associated particle size distribution, are then identified by finding the set of modeled radiances that best approximate the measured radiances in each channel, simultaneously. We calculate expected radiances using measured radiances in channel A5, which was the channel that was used to retrieve the initial $\text{CO}_{\text{2}}$ ice opacity profiles. We did not model radiances in channels A1, A2, and A3 because they are sensitive to $\text{CO}_{\text{2}}$ gas radiance, which is not accounted for in this simple model. Any modeled radiances in these gas channels are therefore systematically lower than the corresponding measured radiances as a result of this significant missing component.

\section{Results--South Polar $\text{CO}_{\text{2}}$ Ice Clouds} \label{sec:Results}

\subsection{Size Distributions} \label{subsec:bestfit}
\begin{figure}[t!]
\centering
\includegraphics[scale=.45]{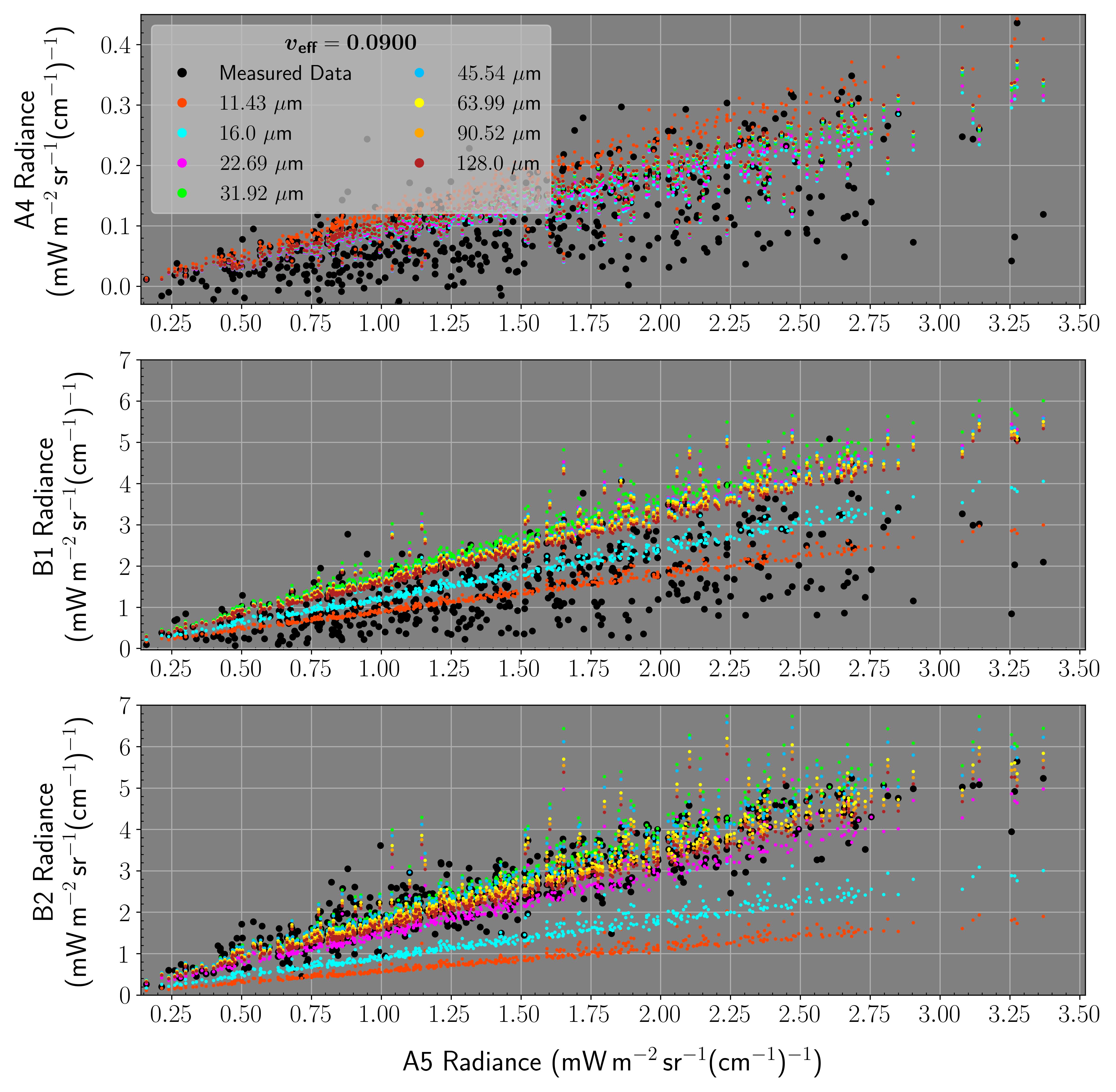}
\caption{Radiance in MCS channels A4, B1, and B2 versus radiance in channel A5. The large black dots represent the set of 507 measured radiances discussed in Section \ref{subsec:bestfit}, while the populations of colored dots represent modeled radiances using Equation \ref{equation:eq19}, for different values of $r_{\text{eff}}$. The effective variance was held constant at a value of $9 \times 10^{-2}$. Note that none of the $r_{\text{eff}}$, $v_{\text{eff}}$ combinations used in this figure were able to fit the measured data sufficiently well in each channel simultaneously, and were therefore excluded from consideration.
\label{fig:fig3}}
\end{figure}

\begin{figure}[t!]
\centering
\includegraphics[scale=.45]{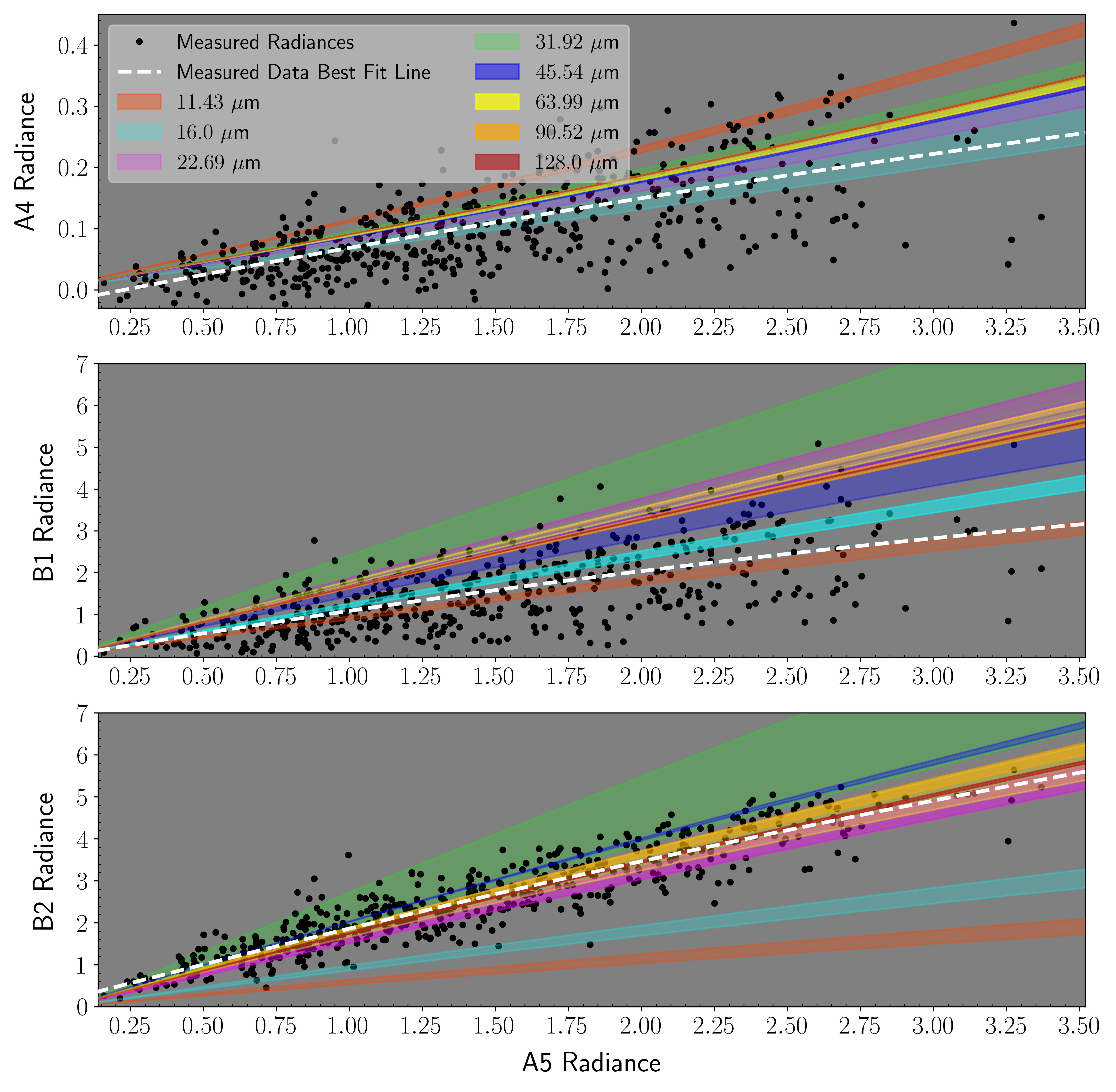}
\caption{Radiance in MCS channels A4, B1, and B2 versus radiance in channel A5. Black dots same as Figure \ref{fig:fig3}. Shaded regions refer to modled radiances, and show the range of radiance space covered by each particle size for the full range of effective variance values considered. White dashed lines represent a second order polynomial fit to the measured data. Radiance units are the same as in Figure \ref{fig:fig3}.
\label{fig:fig4}}
\end{figure}

After reducing our initial dataset following the procedure outlined in Section \ref{subsec:Methods}, we were left with a set of 507 measured radiances in each spectral channel within the south polar region, across 59,625 total observations over the nine months we considered. Equation (\ref{equation:eq19}) was used to model radiances in the three MCS aerosol channels based on the remaining measured radiances in A5, for a range of scattering properties. Scattering properties were calculated with \emph{Supermie}, the Mie code described in \citeA{paige1985annual}, using infrared refractive indices for $\text{CO}_{\text{2}}$ ice from \citeA{hansen_spectral_1997}. The Mie code we used assumes particles are of uniform composition, perfectly spherical, and follow a modified gamma distribution which can be tuned by providing different shape parameters as inputs. Although assuming spherical particles does introduce a source of error, $\text{CO}_{\text{2}}$ has been observed to crystallize into octahedral-/cubo-octahedral-shaped particles under Mars atmosphere-like conditions \cite{MANGAN2017201, FosterChangco2shape}. Because these particles are roughly spherical in shape, we do not anticipate this source of error to be significant, especially compared to other sources of error, such as those generated by uncertainties in the instrument spectral response outlined in Section \ref{subsec:SysErr}.

Because our goal is to determine the particle size distribution that produces Mie parameters that best fit the measured data, each of $r_{\text{eff}}$, $\alpha$, and $\gamma$ are free parameters in our model. Based on \citeA{hayne_carbon_2012}, we imposed an initial constraint of $10 \ \mu m \leq r_{\text{eff}} \leq 100 \ \mu m$, and began by considering seven particle sizes in that range. We also limited $\alpha$ and $\gamma$ to the ranges $1 \leq \alpha \leq 30$, and $0.3 \leq \gamma \leq 20$, in an attempt to reduce the size of our almost entirely unconstrained parameter space. These limits for $\alpha$ and $\gamma$ were chosen because they are broad enough to encompass values used to describe typical terrestrial aerosol distributions \cite{petty_modified_2011, hess_optical_1998, tomasi_infrared_1976, deirmendjian_electromagnetic_1969}.

\begin{figure}[t!]
\centering
\includegraphics[width=.9\textwidth]{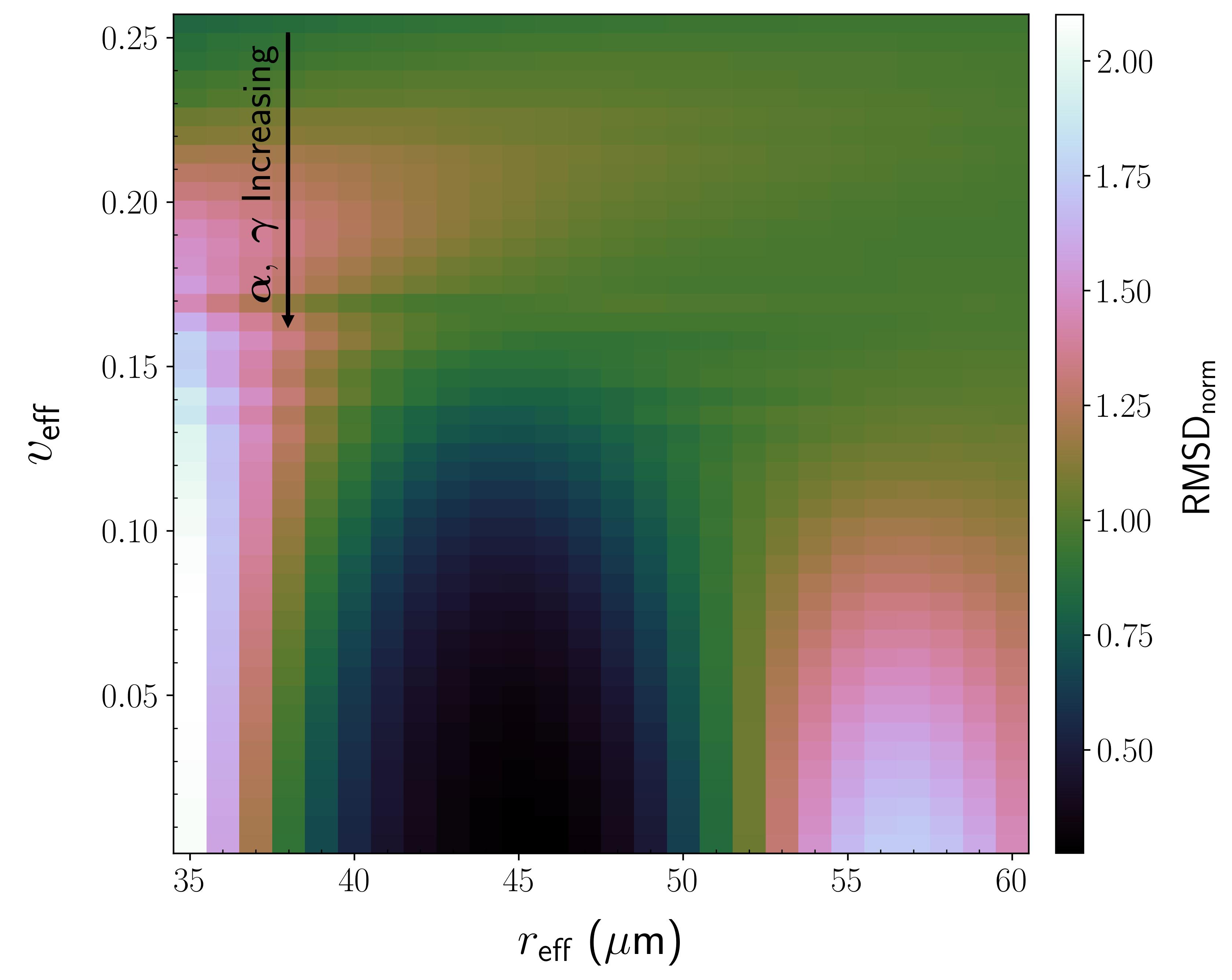}
\caption{RMS deviation versus effective variance and effective radius. Lower (darker) values of $\text{RMSD}_{\text{norm}}$ indicate a better model fit across all aerosol channels. The lowest RMSD value corresponds to a size distribution with $r_{\text{eff}} = 46 \ \mu m$ and $v_{\text{eff}} = 2.0 \times 10^{-3}$.
\label{fig:fig5}}
\end{figure}

Figure \ref{fig:fig3} shows modeled radiances generated for each of the seven effective radii considered, using a particle size distribution with an effective variance of $9 \times 10^{-2}$, plotted against the corresponding measured radiances in channel A5. 
While some parameter combinations do produce radiances that fit the observed measurements well in a single channel, none can accurately fit measured radiances in every channel simultaneously.

\begin{figure}[t!]
\centering
\includegraphics[scale=.45]{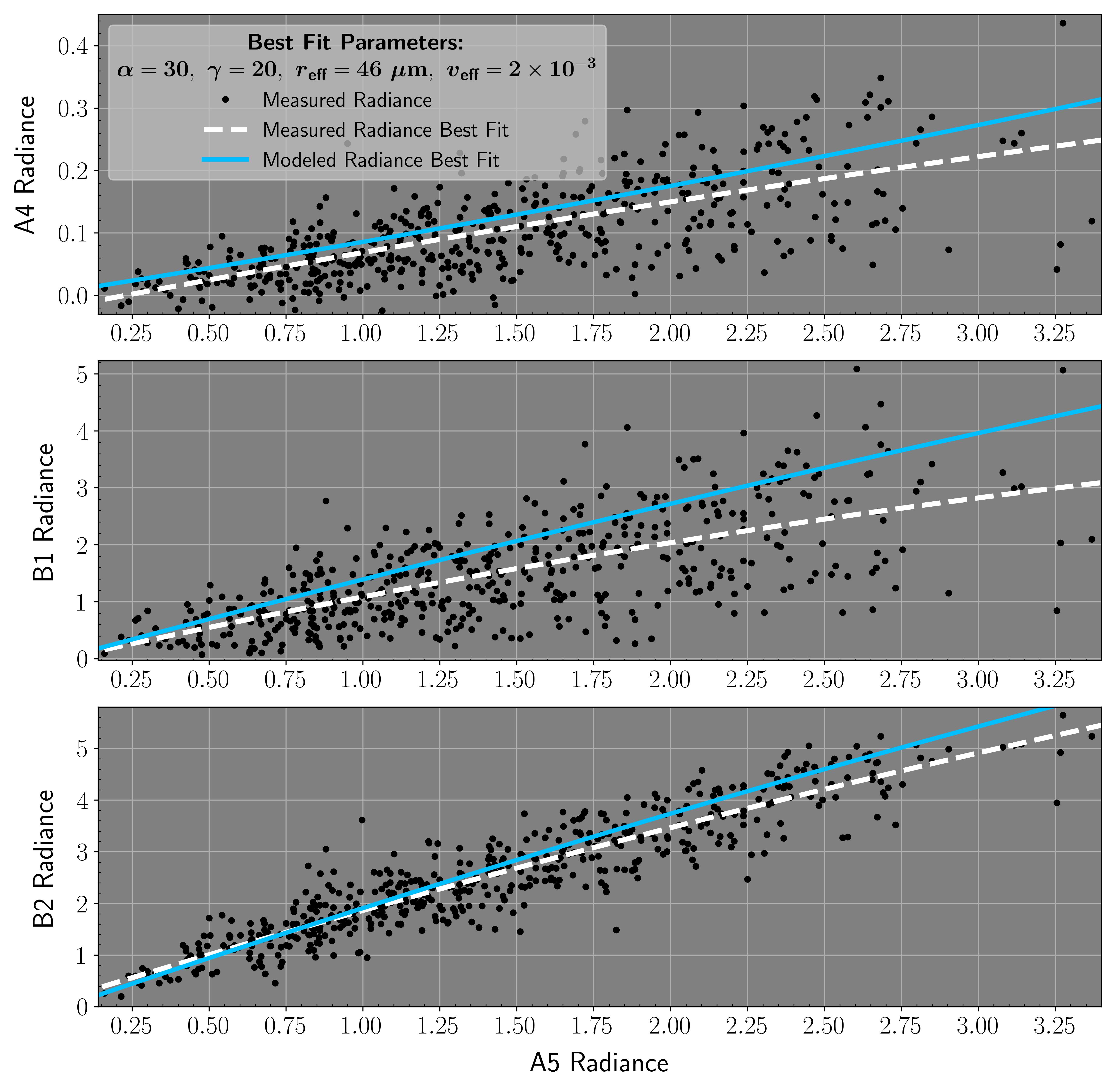}
\caption{Radiance in MCS channels A4, B1, and B2 versus radiance in channel A5. Black dots same as Figure \ref{fig:fig3}. White dashed lines are second-order polynomial fits to the measured data, solid blue lines are second-order polynomial fits to the modeled radiances, generated using our best-fit model parameters. A well-fitting model effectively minimizes the distance between the blue and white lines. Radiance units are the same as in Figure \ref{fig:fig3}.
\label{fig:fig6}}
\end{figure}

To better explore our parameter space, we fit each population of modeled radiances with a second-order polynomial; Figure \ref{fig:fig4} shows the full range of radiance space swept out by these best fit lines for the seven $r_{\text{eff}}$ considered, as $v_{\text{eff}}$ decreases. Because the wings of the modified gamma distribution tend to shrink as both $\alpha$ and $\gamma$ increase, large values of $v_{\text{eff}}$ will correspond to small $\alpha$, $\gamma$ combinations, and vice versa. From Figure \ref{fig:fig4} we see that there are no $\alpha$, $\gamma$ combinations that can adequately fit the data with $r_{\text{eff}} < 32 \ \mu m$ and $r_{\text{eff}} > 64 \ \mu m$, allowing us to further reduce our parameter space.

Although particle sizes in the range $32 \ \mu m < r_{\text{eff}} < 64 \ \mu m$ produce distributions that can fit the measured data in each channel independently, valid particle size distributions are only those that can fit the measured data in each channel simultaneously. To investigate how $\alpha$ and $\gamma$ influence the goodness of our model’s fit, we calculate the sum of the normalized root mean square deviation (RMSD) in each spectral channel. For a given parameter combination, $i$, the normalized RMSD, $RMSD_{\text{norm}_{i}}$, is given by
\begin{equation} \label{equation:eq21}
    RMSD_{\text{norm}_{i, \, j}} = \displaystyle \sum_{j = 0}^{M} \frac{RMSD_{i, \ j} - RMSD_{\text{min}_{i, \, j}}}{RMSD_{\text{max}_{i, \, j}} - RMSD_{\text{min}_{i, \, j}}}
\end{equation}
\noindent where $M = 4$ is the total number of spectral channels, $RMSD_{\text{min}_{i, \, j}}$ is the minimum RMSD in channel $j$, $RMSD_{\text{max}_{i, \, j}}$ is the maximum RMSD in channel $j$, and $RMSD_{i, \ j}$ is the root mean square deviation in channel $j$, given by
\begin{equation} \label{equation:eqrmsd}
    RMSD_{i, \ j} = \sqrt{\frac{\sum_{k = 1}^{N} ( x_k - \hat{x}_k )^2}{N}}
\end{equation}
\noindent where $\{ x_1, \ x_2, \ ..., \ x_N \}$ are measured radiances, $\{ \hat{x}_1, \ \hat{x}_2, \ ..., \ \hat{x}_N \}$ are corresponding calculated radiances, and $N$ is the total number of radiances considered.

Figure \ref{fig:fig5} shows this RMSD statistic for different values of $r_{\text{eff}}$ and $v_{\text{eff}}$, with lower values of RMSD indicating a better fit to the measured data. We see that the best fit between our model and the measured data occurs for values of $r_{\text{eff}} \approx 46 \ \mu m$, and $v_{\text{eff}} \leq 2.0 \times 10^{-3}$, corresponding to the most narrow particle size distributions considered.

\begin{wrapfigure}{r}{0.62\linewidth}
\includegraphics[scale=.5]{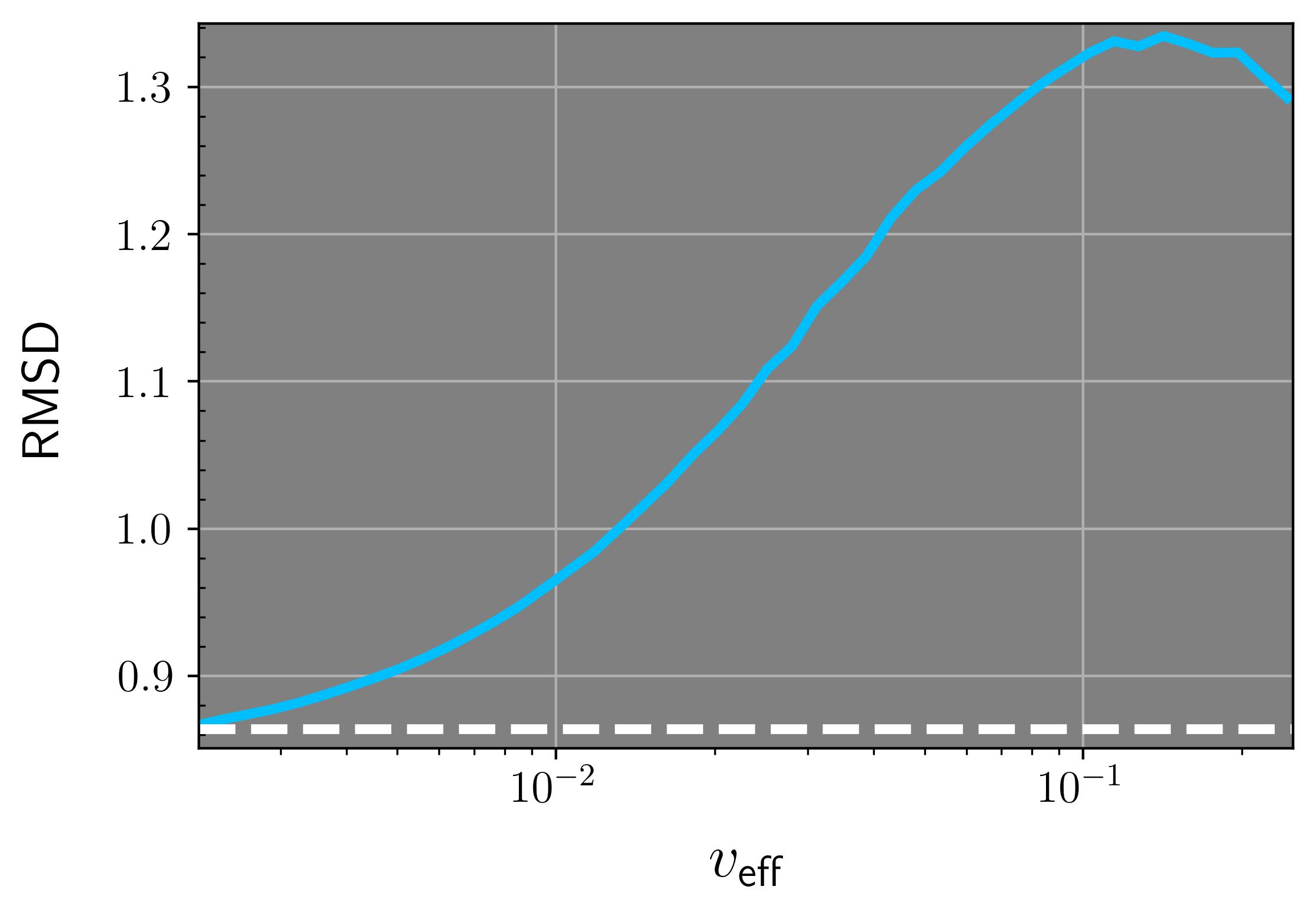}
\caption{RMSD vs $v_{\text{eff}}$ in channel B1 for our best-fit model. The solid white line represents the horizontal asymptote of $\sim 0.85$, reached as $v_{\text{eff}}$ decreases.
\label{fig:fig7}}
\end{wrapfigure}

Figure \ref{fig:fig6} compares radiances generated using Mie parameters from the best-fit size distribution with measured data in each channel. Qualitatively, the fit is good in channels A4 and B2, but modeled radiances in B1 are slightly too high. A better fit can be achieved in B1 at the expense of the goodness of fit in the other channels, especially A4. This misfit in B1 may be due to a number of different factors: it may indicate that our chosen parameter space was too restrictive, that we have neglected important physical processes, or it may be the result of systematic uncertainties in B1 measurements. The first possibility listed here would be the simplest to address; however, the RMS deviation in channel B1 versus $v_{\text{eff}}$ for our best-fit particle size of 46 $\mu$m (Figure \ref{fig:fig7}) shows that the RMSD asymptotes to a value of around 0.86 as $v_{\text{eff}}$ decreases. This indicates that extending our parameter space to larger values of $\alpha$ and $\gamma$ (smaller values of $v_{\text{eff}}$) will not provide a better model fit, eliminating this first possibility.

\subsection{Effects of Aerosol Mixtures} \label{subsec:AerMix}
\subsubsection{Water Ice Condensation Nuclei} \label{subsubsec:H2OCN}

Water has a much higher frost point than $\text{CO}_{\text{2}}$ which causes most of the water vapor present in the Martian atmosphere to freeze out before $\text{CO}_{\text{2}}$ ice condensation begins. If water ice particles do not grow large enough to sediment out of the atmosphere before $\text{CO}_{\text{2}}$ begins to freeze, they can serve as $\text{CO}_{\text{2}}$ ice condensation nuclei along with suspended dust. Based on these physical considerations, a subset of $\text{CO}_{\text{2}}$ ice particles in the Martian atmosphere are therefore expected to be composed of a shell of $\text{CO}_{\text{2}}$ ice surrounding a small water ice core.

While $\text{CO}_{\text{2}}$ ice is relatively opaque in much of the far and mid infrared, channels A5 and B1 measure emission within a significant transparency band in the $\text{CO}_{\text{2}}$ ice spectrum, centered around 25 $\mu$m. Radiance measurements in these channels may therefore contain a detectable signature from the water ice core, despite the fact that we have eliminated profiles with significant retrieved water ice opacity. If this signature is large enough, it could provide an explanation for the slight misfit in B1.

Mixtures of $\text{CO}_{\text{2}}$ and $\text{H}_{\text{2}}\text{O}$ ices have been observed in a limited number of mesospheric $\text{CO}_{\text{2}}$ ice clouds by the Atmospheric Chemistry Suite onboard the Trace Gas Orbiter \cite{LUGININ2024116271}. While the authors of the aforementioned study did not explore the exact nature of these mixtures, their observations could be explained by composite $\text{CO}_{\text{2}}/\text{H}_{\text{2}}\text{O}$ ice particles. Given that these internal mixtures may exist within mesospheric $\text{CO}_{\text{2}}$ ice clouds, it is likely that they may also be present within $\text{CO}_{\text{2}}$ ice clouds in the polar regions where water vapor transport is enhanced during the winter months.

To quantify the effect of spectral contamination by water ice CN, we approximate clouds composed of these $\text{CO}_{\text{2}}$ ice/$\text{H}_{\text{2}}\text{O}$ ice hybrid particles as an external mixture of two independent size distributions of $\text{CO}_{\text{2}}$ ice and water ice with identical number densities. Following \citeA{kleinbohl_single-scattering_2011}, we use water ice size distributions with $\alpha = 2$, $\gamma = 1.5$, and $r_{\text{eff}}$ between 0.25 $\mu$m and 16.0 $\mu$m. We then calculate a set of effective extinction efficiencies for the mixture using optical constants from \citeA{warren_optical_1984} and Equation \ref{equation:eq13}. For simplicity, modeled radiances in this section and in Section \ref{subsubsec:AltGrad} neglect the wavelength dependence of $\varpi_0$ and $\bar{P}$ such that Equation \ref{equation:eq19} reduces to 
\begin{equation} \label{equation:eqnoscat}
    R_{\text{e}_{\text{ch2}}} = \frac{B_{\nu_{\text{ch2}}} \left( T_{\text{eff}} \right) \, Q_{\text{ext}_{\text{ch2}}}}{B_{\nu_{\text{ch1}}} \left( T_{\text{eff}} \right) \, Q_{\text{ext}_{\text{ch1}}}} R_{\text{m}_{\text{ch1}}}
\end{equation}

\begin{figure}[t!]
\includegraphics[width=.9\textwidth]{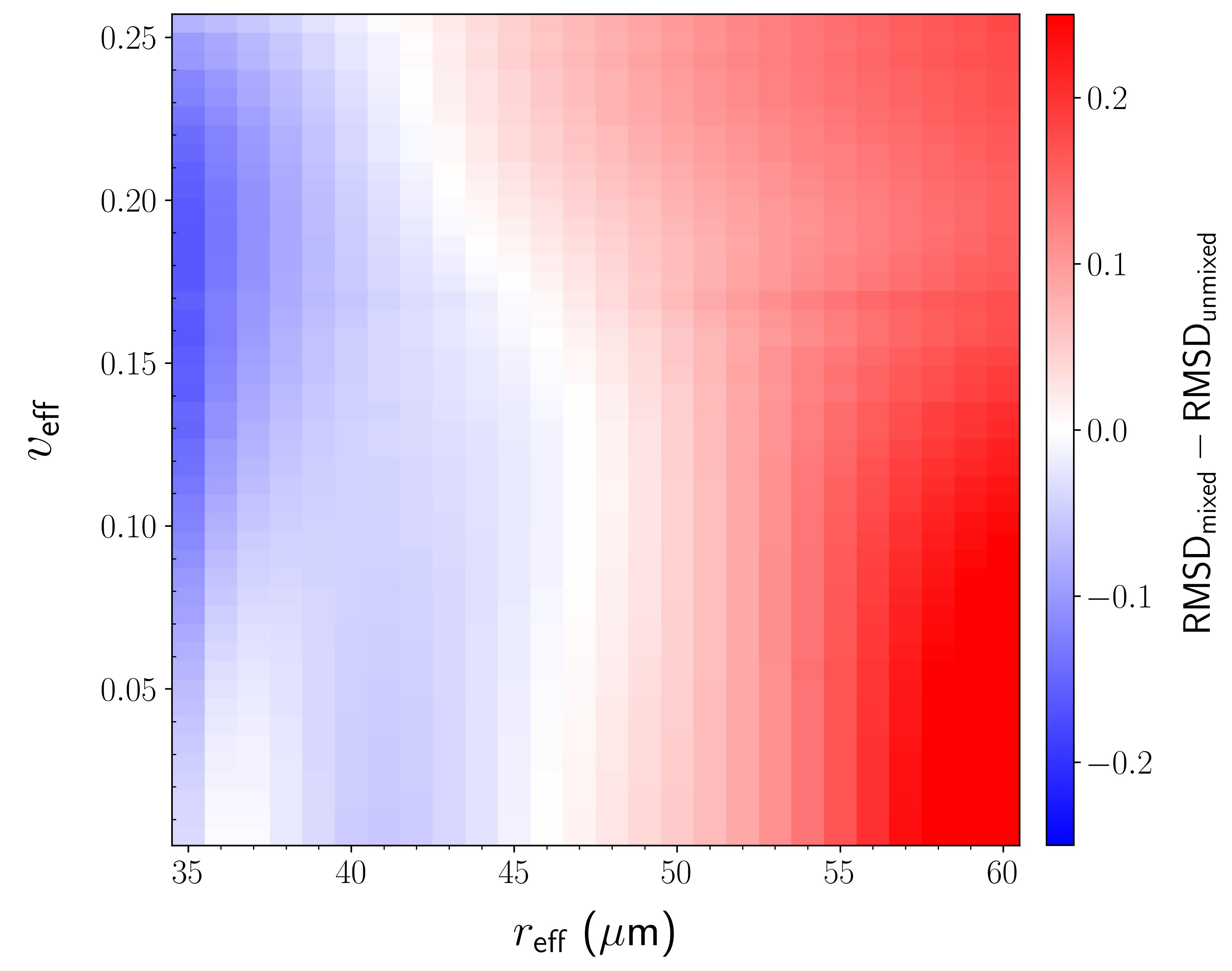}
\caption{Difference in normalized RMSD between the pure $\text{CO}_{\text{2}}$ ice and $\text{CO}_{\text{2}}$ ice + $\text{H}_{\text{2}}\text{O}$ ice mixture as a function of $\text{CO}_{\text{2}}$ ice particle size and effective variance. The water ice/$\text{CO}_{\text{2}}$ ice mixture uses the best-fit $\text{H}_{\text{2}}\text{O}$ ice particle size distribution which has an effective radius of 11.3 $\mu$m and an effective variance of 0.15. Note that values are very small, especially near the best-fit $\text{CO}_{\text{2}}$ ice particle size of 46 $\mu$m, indicating that $\text{H}_{\text{2}}\text{O}$ ice CN have little effect.
\label{fig:fig8}}
\end{figure}

We take this approach because $\bar{P}$ must be calculated for each observed radiance in each spectral channel for each parameter combination, something that becomes computationally expensive for aerosol mixtures, which themselves increase the number of free parameters that must be fit. We can estimate the effect that this simplification will have on our final results by repeating the analysis in Section \ref{subsec:bestfit} using Equation \ref{equation:eqnoscat} instead of Equation \ref{equation:eq19}. Taking this approach ultimately did not change our best-fit values of $r_{\text{eff}}$ and $v_{\text{eff}}$ from what was obtained using the more accurate method. We therefore expect that using this simplified approach for aerosol mixtures will have little impact on our final results.

\begin{figure}[b!]
\centering
\includegraphics[scale=.45]{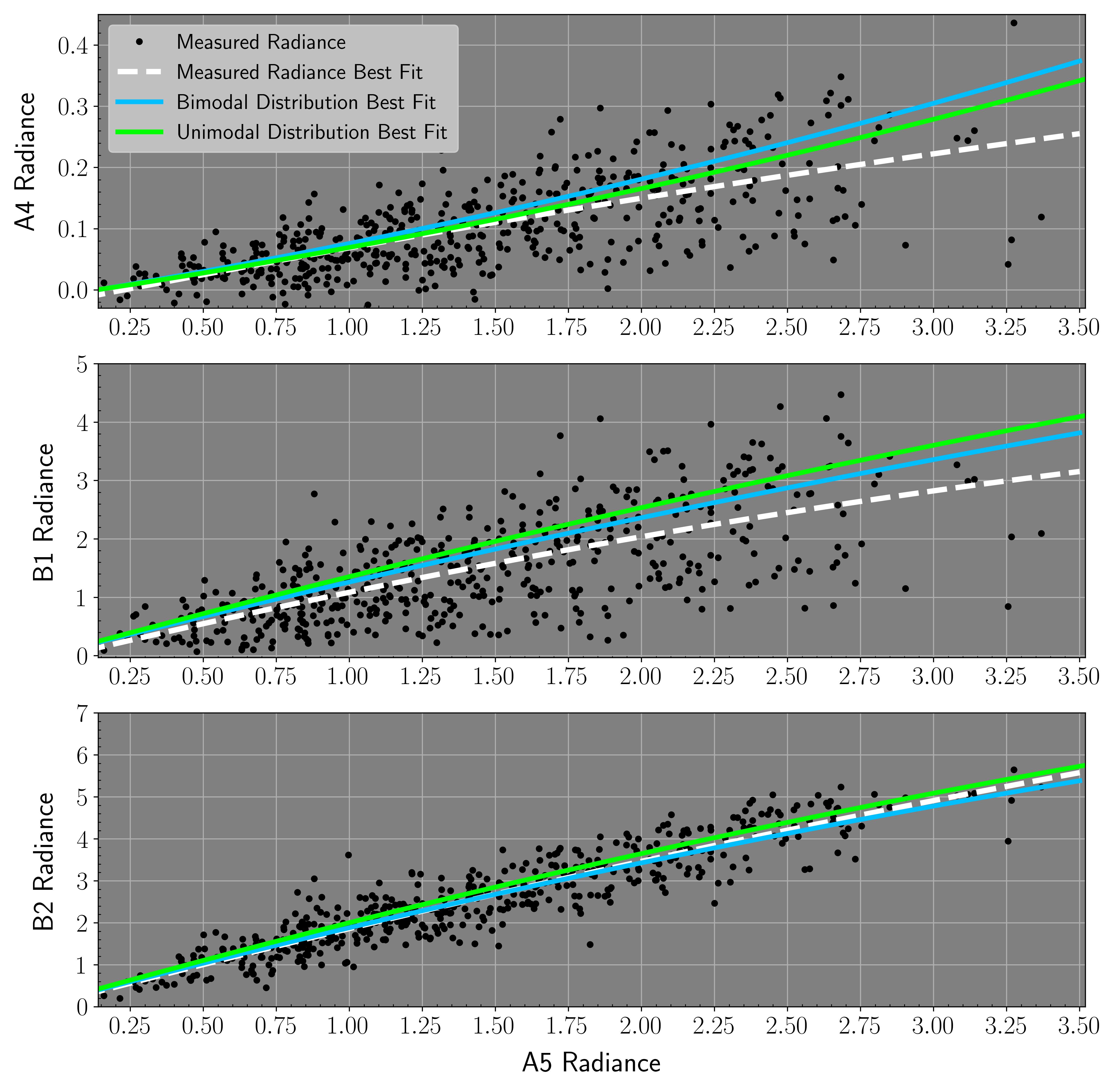}
\caption{Radiance in MCS channels A4, B1, and B2 versus radiance in channel A5. The black dots are the same as in Figure \ref{fig:fig3}. The white dashed lines are second-order polynomial fits to the measured data, and the blue and green lines correspond to second-order fits to the modeled radiances using our best-fit bimodal and unimodal particle size distributions, respectively. The unimodal size distribution is the same as the one used in Figure \ref{fig:fig6}. The best-fit bimodal distribution is made up of two component size distributions, the first of which has $r_{\text{eff}} = 9.25 \ \mu$m and $v_{\text{eff}} = 0.15$, while the second has $r_{\text{eff}} = 44 \ \mu$m and $v_{\text{eff}} = 2.0 \times 10^{-3}$. The relative number density of particles in the smaller distribution to the larger one is 4:1. Radiance units are the same as in Figure \ref{fig:fig3}.
\label{fig:fig9}}
\end{figure}

For water ice cores with $r_{\text{eff}} \leq 8 \ \mu m$, our results are nearly identical to the pure-CO$_2$ ice case. This is due to the fact that the extinction efficiencies of water ice and $\text{CO}_{\text{2}}$ ice are weighted by the square of their respective particle size in Equation \ref{equation:eq13}. With equal number densities, the final mixed extinction efficiencies are dominated by $\text{CO}_{\text{2}}$ ice due to its much larger particle size. Figure \ref{fig:fig8} shows the difference in RMSD between the pure $\text{CO}_{\text{2}}$ ice, and $\text{CO}_{\text{2}}$ + $\text{H}_{\text{2}}\text{O}$ ice scenarios for the best-fit water ice particle size of 11.3 $\mu$m. While the fit is marginally improved for the water ice mixture, the difference in minimum RMSD between the two scenarios is only $\sim 1\%$. In addition, water ice particles rarely grow this large at altitudes above $\sim$10 km on Mars due to its thin and arid atmosphere, making such a mixture physically unlikely \cite{fedorova_solar_2009, clancy_mars_2003}. We therefore do not anticipate contamination from water ice CN cores to be a major source of uncertainty in our derived best-fit properties.

\subsubsection{Vertical Gradients in Particle Size} \label{subsubsec:AltGrad}

Atmospheric models of the Martian polar regions have shown that $\text{CO}_{\text{2}}$ ice particle size should decrease with altitude \cite{maattanen_troposphere--mesosphere_2022, colaprete_formation_2003}. Because different MCS detectors sample from different atmospheric layers, for observations of extended clouds MCS measurements will capture radiance produced by $\text{CO}_{\text{2}}$ ice particles with a variety of different particle sizes. While the MCS retrieval algorithm does not support altitude-dependent particle size distributions, their effects on measured radiances may be significant, and should be investigated.

We approach this problem by considering a layer of fine-grained $\text{CO}_{\text{2}}$ ice overlaying a cloud deck consisting of much larger particles. We approximate this by using Equation (\ref{equation:eq13}) to mix $\text{CO}_{\text{2}}$ ice Mie parameters from two different size distributions; one small, with an effective radius between 10-20 $\mu$m, and another larger distribution with $r_{\text{eff}}$ between 40-50 $\mu$m. This creates a bimodal size distribution with two peaks in radius. The addition of this second particle size distribution more than doubles the number of free parameters in our model, as the final result now depends on both a second set of unconstrained shape parameters, as well as the number densities of small and large particles.

We find that the normalized RMSD statistic is minimized for a relative number density of $\sim 4:1$ small to large particles. While the minimum RMSD for a bimodal size distribution with this number density ratio is lower than for an unmixed, unimodal size distribution, the difference is within our observational uncertainty (Figure \ref{fig:fig9}). We therefore cannot rule out the possibility that small $\text{CO}_{\text{2}}$ ice particles have some influence on radiances measured by MCS. However, the magnitude of this effect is small compared to the impact that the larger 46 $\mu$m particles have. The insensitivity of our results to the presence of smaller particles at higher altitudes indicates there is little justification for including this complication in the MCS retrieval algorithm.

\subsection{Potential Systematic Errors} \label{subsec:SysErr}

\begin{wrapfigure}{r}{0.62\linewidth}
\includegraphics[scale=.5]{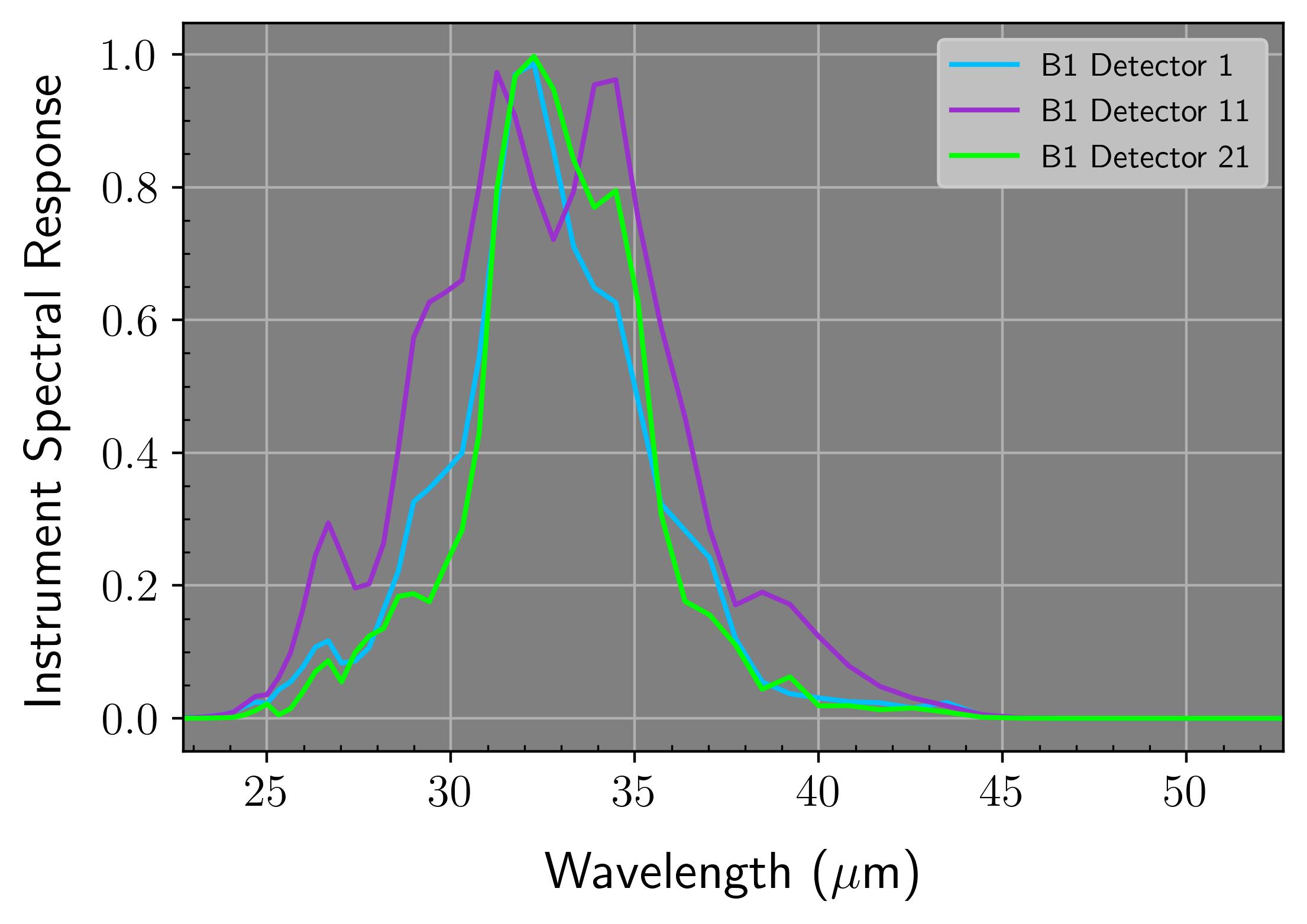}
\caption{Instrument spectral response in B1 for the three detectors characterized before launch.
\label{fig:fig10}}
\end{wrapfigure}

In this section, we describe potential systematic errors that could provide an explanation for the model misfit in B1 (Figure \ref{fig:fig6}). The most likely of these lies in the spectral response of the B channel detectors. Unlike the A channels, the MCS B channels have significant detector-to-detector variability in their spectral response due to the fact that different filter types were used between the two channels. Unfortunately, this rather significant variability was not fully characterized in channel B1 before launch due to time constraints. Only three detectors in B1 (1, 11, and 21) were able to be measured; the spectral response of the other detectors was generated by interpolating between the known three.

Uncertainties in the B-channel spectral response could measurably effect modeled radiances given that both the extinction efficiencies, and Planck radiances in Equation \ref{equation:eq19} are band-averaged quantities. Because the B channels are not frequently used by the MCS retrieval algorithm in its present form, errors introduced by this interpolation are typically small. However, as can be seen in Figure \ref{fig:fig10}, the shape of the B1 detector response function varies significantly among the three that are known. Because of this, the interpolation method used to generate the unknown detector spectral responses is unlikely to fully capture their true behavior; however, without a targeted observational campaign, the extent of this uncertainty cannot be fully understood. This detector-to-detector variability therefore presents a plausible explanation that can account for the B1 misfit observed in Figure \ref{fig:fig6}, especially given that the misfit itself is small compared to the scatter in the measured radiances.

\subsection{Why is the Derived Particle Size Distribution so Narrow?} \label{subsec:Narrow}


Particle size distributions are convenient statistical representations of ensembles of particles, but they are not without physical significance as well; the underlying microphysical processes occurring during ice condensation influence the distribution such that its shape provides information about local atmospheric properties and dynamics. During active $\text{CO}_{\text{2}}$ ice cloud formation, the two primary mechanisms that shape particle size distributions are particle growth and sedimentation. The growth timescale for a spherical particle to accumulate a layer of ice of thickness $dr$ scales as $r^2$, where $r$ is the particle radius. Therefore, as the particle grows and increases its surface area, it takes longer and longer to accumulate a layer of ice of constant thickness. This strong dependence on particle size means that smaller particles are able to ``catch up'' in size with larger ones due to their much more rapid growth rates, effectively removing smaller particles from the size distribution.

On the other end of the size distribution, larger particles sediment out of the atmosphere more quickly than smaller ones. The characteristic timescale for the removal of particles from the atmosphere by sedimentation is on the order of the gravitational settling time, or the time it takes for a particle to freely fall a single scale height. For spherical particles, the settling time can be written:

\begin{equation} \label{equation:eqtset}
    t_s \left( r \right) = \frac{H}{w_0 \left( r \right)}
\end{equation}

\noindent where $H$ is the atmospheric scale height, and $w_0 \left( r \right)$ is the gravitational settling velocity provided by the Stokes-Cunningham law for spherical particles

\begin{equation} \label{equation:eqwset}
    w_0 \left( r \right) = \left( \frac{2 \rho_s g r^2}{9 \eta} \right) \left( 1 + \frac{\lambda}{r} \right)
\end{equation}

\noindent where $\rho_s$ is the solid density of the particle, $g$ is gravitational acceleration, $r$ is the particle radius, $\eta$ is the molecular viscosity of the atmosphere, and $\lambda$ is the molecular mean free path \cite{hayne_role_2014}. From Equations \ref{equation:eqtset} and \ref{equation:eqwset}, we can see that the characteristic settling time also decreases as particles grow larger. In the lowermost atmospheric scale height, $\frac{\lambda}{r} \ll 1$ and $t_\mathrm{s} \sim r^{-2}$, such that growth in $r$ by a factor of 2 leads to $4\times$ faster settling. This means that once particles grow large enough to sediment out of the atmosphere, larger particles are removed from the size distribution much more efficiently than smaller ones.

Together, these two removal processes control the spread in the $\text{CO}_{\text{2}}$ ice size distribution by limiting the minimum and maximum particle sizes that can be present. Depending on how quickly these processes act in the Martian atmosphere, this may present a possible physical mechanism for forming the extremely narrow size distributions we observe in the MCS data. To test this hypothesis, we construct a simple 1D convective cooling model to simulate the growth and sedimentation of $\text{CO}_{\text{2}}$ ice driven by convective storms in the Martian polar regions.

\subsubsection{1D Convective Cooling Model} \label{subsubsec:CoolingModel}

\citeA{colaprete_formation_2003} showed that convective storms may be significant drivers of $\text{CO}_{\text{2}}$ snowfall on Mars. Our model simulates the effects of one of these convective storms by balancing the total cooling in each atmospheric layer with the release of latent heat as $\text{CO}_{\text{2}}$ ice forms to calculate a mass growth rate given by

\begin{equation} \label{equation:eqco2grow}
    \frac{dm}{dt} = -\frac{\Delta P c_p}{g L N dz} \left( \frac{dT}{dt} \right)
\end{equation}

\noindent where $\frac{dm}{dt}$ is the mass growth rate of $\text{CO}_{\text{2}}$ ice, $\Delta P$ is the pressure difference between the top and bottom boundaries of an atmospheric layer, $c_p$ is the specific heat of $\text{CO}_{\text{2}}$, $g$ is gravitational acceleration, $L$ is the latent heat of sublimation of $\text{CO}_{\text{2}}$, $N$ is the number density of $\text{CO}_{\text{2}}$ ice particles in a layer, $dz=0.1$ km, is the layer thickness, and $\frac{dT}{dt}$ is the convective cooling rate:
\begin{equation} \label{equation:eqconvcool}
    \frac{dT}{dt} = -w_z \Gamma
\end{equation}
\noindent where $w_z=1.5$ m/s, is the vertical wind speed--consistent with the average updraft velocity at the south pole from \citeA{colaprete_formation_2003}, and $\Gamma= g/c_\mathrm{p} = 4.5$ K/km is the adiabatic lapse rate on Mars. For simplicity we do not model the full microphysics of $\text{CO}_{\text{2}}$ condensation, which is beyond the scope of this paper.

\begin{figure}[b!]
\centering
\includegraphics[width=.9\textwidth]{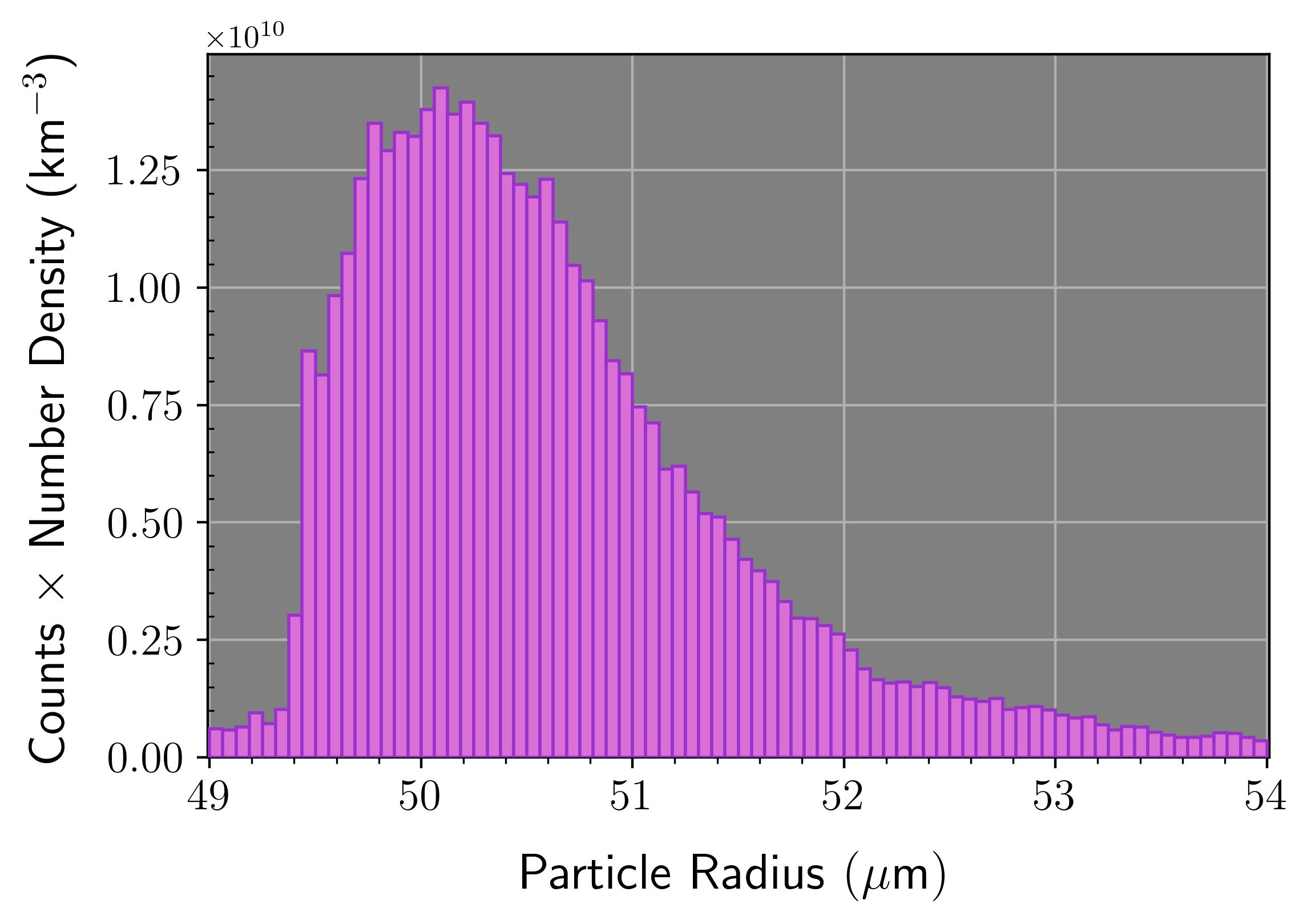}
\caption{Histogram of simulated particle size generated using our convective cooling model described in Section \ref{subsubsec:CoolingModel}. The shape of the distribution follows a modified gamma distribution with $r_{\text{eff}} = 50.9$ $\mu$m and $v_{\text{eff}} = 1.0 \times 10^{-3}$.
\label{fig:fig11}}
\end{figure}

We track the motion of $\text{CO}_{\text{2}}$ ice particles in our model by first using Equations \ref{equation:eqco2grow} and \ref{equation:eqconvcool} to calculate how much each particle has grown after each time step, after which Equation \ref{equation:eqwset} is used to update settling velocities. Particles are assumed to move upward at the vertical wind speed, $w_z$, so their net vertical velocity is given by $w_{\text{net}} = w_z + w_0 \left( r \right)$, where $w_0 \left( r \right)$ is always negative. An initial size distribution consistent with the one used by \citeA{kleinbohl_mars_2009} for dust is used to simulate $\text{CO}_{\text{2}}$ ice CN. Particles begin in the lowermost model layer and are allowed to move vertically depending on their respective net velocities. To simulate replenishment by horizontal advection, new condensation nuclei are continually added to the bottom of the atmosphere such that the number density of the lowermost layer never changes. The initial particle number density, $N$, is a tuneable free parameter which is set to 5.0$\times 10^{14}$ km$^{-3}$, a value falling between the upper limits estimated in \citeA{alsaeed_transport_2022} and \citeA{colaprete_formation_2003}. A background number density of $2 \times 10^{14}$ km$^{-3}$ is assigned to atmospheric layers where the number density is too low, in order to limit the runaway growth of particles in those layers. The model is run until particles begin to turn around after reaching their maximum altitude, at which point the size and size distribution of particles above 10 km, which MCS is able to observe, are reported.


\subsubsection{Simulated Particle Size Distributions} \label{subusbsec:simdists}

A histogram of particle size for particles above 10 km is shown in Figure \ref{fig:fig11}. Our simulated distribution follows the general shape of a modified gamma distribution with more extended wings. Using Equations \ref{equation:eq10} and \ref{equation:eq12}, the effective radius and effective variance of this distribution are 50.9 $\mu$m and $1.0 \times 10^{-3}$, respectively. These values are in good agreement with the ones derived from MCS data above, indicating that our proposed mechanism for forming strongly peaked $\text{CO}_{\text{2}}$ ice size distributions has a physical basis.

This may not be entirely unexpected given that the Martian atmosphere is both 95\% $\text{CO}_{\text{2}}$, and much more tenuous than our own. The fact that $\text{CO}_{\text{2}}$ is the most abundant atmospheric constituent means that the growth of $\text{CO}_{\text{2}}$ ice particles is not diffusion limited, allowing particles to grow from $< 1$ $\mu$m to $\sim 10$ $\mu$m within a few seconds in our model. The thin Martian atmosphere also limits how large aerosols can be before rapidly sedimenting out of the atmosphere. While our own atmosphere can support much larger particles for extended periods of time, aerosols on Mars rapidly fall to the surface once they reach sizes of only a few tens of microns. $\text{CO}_{\text{2}}$ ice particles on Mars that are observable by MCS are therefore limited to a very narrow size range.

\section{Results--North Polar $\text{CO}_{\text{2}}$ Ice Clouds} \label{sec:NPco2}

Differences in $\text{CO}_{\text{2}}$ ice particle size between the northern and southern hemispheres have not yet been investigated in detail. The pressure dependence in Equation \ref{equation:eqco2grow} would imply that $\text{CO}_{\text{2}}$ ice should grow larger in the northern hemisphere due to its higher average surface pressures. However, the northern polar regions are also known to have considerably more dust and water ice (and therefore CN) during the winter months than the southern polar regions at wintertime \cite{garybicas_asymmetries_2020}. As the number of CN increases, the total condensable mass of $\text{CO}_{\text{2}}$ is distributed across a larger effective surface area, limiting the rate at which $\text{CO}_{\text{2}}$ ice crystals can grow. The interaction between these two competing processes has yet to be fully understood, with different atmospheric models producing different particle size estimates \cite{maattanen_troposphere--mesosphere_2022, colaprete_co2_2008}.

The fact that $\text{CO}_{\text{2}}$ ice particles in the northern hemisphere may have different physical characteristics than those in the south could introduce a significant source of uncertainty in future $\text{CO}_{\text{2}}$ ice retrievals, if not accounted for. Below, we repeat our particle size study outlined in Section \ref{sec:D&M} for the northern hemisphere, to determine whether or not these differences need to be considered in the MCS retrieval algorithm.

\subsection{Particle Size} \label{subsec:Npsize}

In our analysis of the south pole we grouped radiance profiles into families as part of our data reduction methodology. Unfortunately, the number of usable CO$_2$ cloud profiles in the north is greatly reduced by the much higher water ice abundance during northern winter. As a result, we were required to skip this initial step to maximize the number of usable $\text{CO}_{\text{2}}$ ice profiles available to us. Even after skipping the family analysis, we were only left with a set of 133 usable radiances within the north polar region from the same set of 59,625 total observations.

\begin{figure}[t!]
\centering
\includegraphics[width=.9\textwidth]{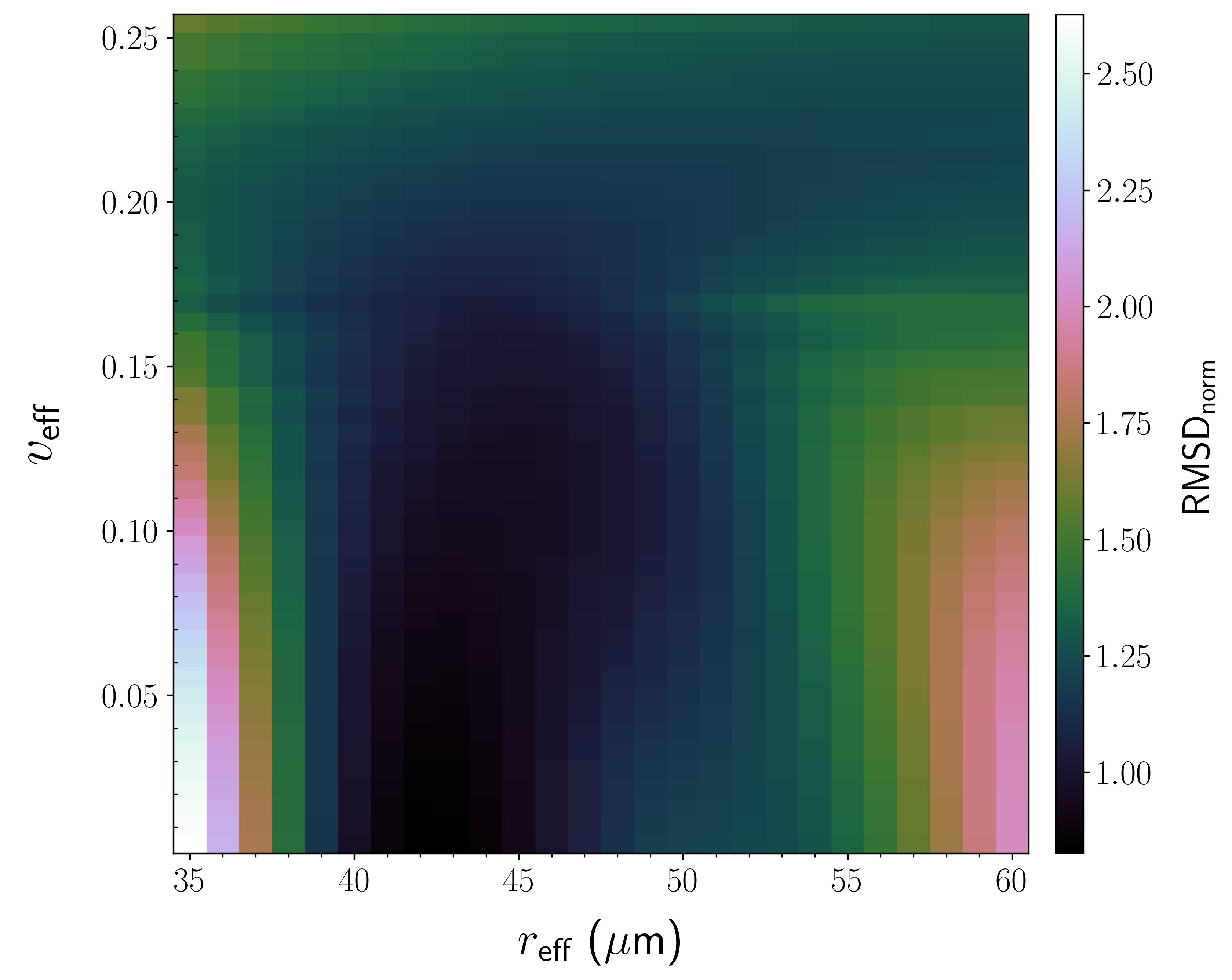}
\caption{Same as Figure \ref{fig:fig5}, this time for the north pole of Mars.
\label{fig:fig12}}
\end{figure}

Figure \ref{fig:fig12} presents the normalized RMSD for different $\text{CO}_{\text{2}}$ ice size distributions in the northern hemisphere. The nearly identical results to those presented for the southern hemisphere suggest that any differences in particle size are difficult for MCS to resolve, and are therefore unlikely to influence future $\text{CO}_{\text{2}}$ ice retrievals. Compared to the southern hemisphere, the best-fit distribution does show larger values of normalized RMSD ($\sim 1.0$ in the north versus $\sim 0.25$ in the south). This is mainly due to the fact that the model struggles to provide an adequate fit in channel A4, where modeled radiances are lower than observed radiances by a factor of almost 2.

\begin{figure}[t!]
\centering
\includegraphics[scale=.45]{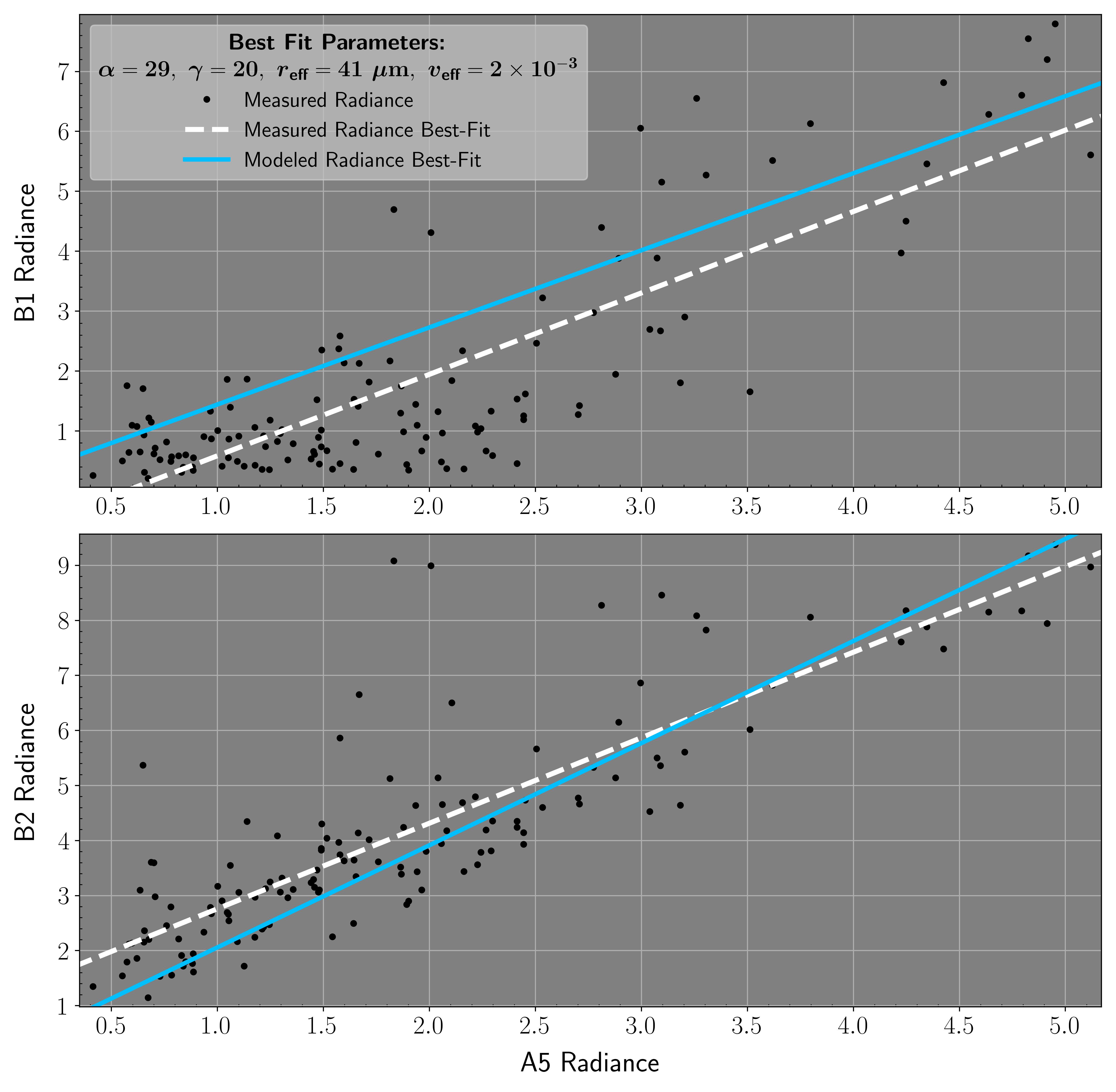}
\caption{Same as Figure \ref{fig:fig6}, also for the north pole. Radiance units are the same as in Figure \ref{fig:fig3}.
\label{fig:fig13}}
\end{figure}

The most likely explanation for the misfit follows from our assumption in Equation \ref{equation:eq19} that all the radiance in the modeled channel can be accounted for by the measured radiance in the reference channel. For example, we cannot use Equation \ref{equation:eq19} to model radiances in the MCS 15-$\mu$m gas channels (i.e. A1, A2, A3) because our reference channel (A5) is not sensitive to $\text{CO}_{\text{2}}$ gas emission. In the southern hemisphere, our assumption that all the observed radiance in A4 is produced only by atmospheric $\text{CO}_{\text{2}}$ ice is valid assuming that we eliminate observations with significant water ice opacity, especially given the lack of atmospheric dust near the south pole during the winter months \cite{hayne_carbon_2012}. This assumption may break down in the northern hemisphere, where the winter atmosphere is characterized by substantially higher dust loading near the pole \cite{montabone2020martian}.

In the dataset used in this study, MCS retrievals of dust opacity are replaced with $\text{CO}_{\text{2}}$ ice opacity, limiting our ability to simultaneously determine how much dust and CO$_2$ ice are present. Because of their shorter central wavelengths, channels A4 and A5 are more sensitive to dust than B1 or B2, meaning that observed radiances in A4 and A5 could contain an extra radiance component from dust that would not be measured in the other aerosol channels. If this is the case, the model would be unable to accurately approximate A4 radiances in the north polar region where dust loading is high. We therefore exclude A4 from our final analysis, focusing instead on B1 and B2.

Modeled radiances using our best-fit size distribution with A4 excluded are shown in Figure \ref{fig:fig13}. The model approximates the data quite well in the two relevant aerosol channels, despite not producing a good fit in A4. From this analysis we derive a best-fit size distribution very similar to the one for the south, with $r_{\text{eff}} = 42$ $\mu$m, and $v_{\text{eff}} = 2.0 \times 10^{-3}$. 

It may be unexpected that derived particle sizes are so similar in the two hemispheres; however, upon inspection of the retrieved atmospheric profiles, we find that $\text{CO}_{\text{2}}$ ice clouds tend to form around the same pressure levels in both hemispheres. This means that $\text{CO}_{\text{2}}$ ice particles near the top of these clouds should rapidly grow to similar sizes before falling out of the atmosphere. Particle sizes in the northern hemisphere may well exceed those in the southern hemisphere near the surface. However, the optical depth at these altitudes is generally quite high, which limits the ability of MCS to provide information regarding these particles (especially below $\sim 5$ km). Alternatively, the extra dust and water ice in the northern hemisphere may limit the maximum size of $\text{CO}_{\text{2}}$ ice particles by increasing the number of available condensation nuclei and distributing the total mass of condensable $\text{CO}_{\text{2}}$ across a larger effective surface area.

\begin{figure}[t!]
    \centering
    \includegraphics[width=.9\textwidth]{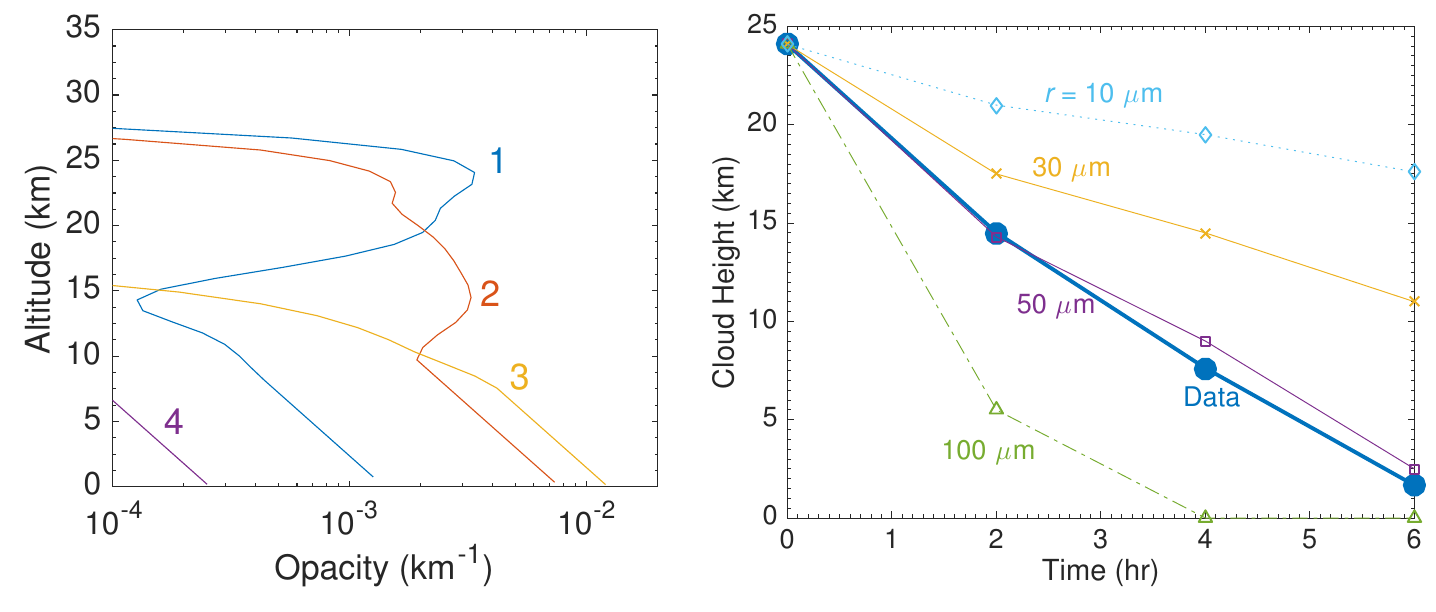}
    \caption{(Left) North polar cloud opacity profiles retrieved from MCS targeted observations during the winter of MY 32 (December 2014) at $L_\mathrm{s} = 246^\circ$ on four sequential orbital passes (numbered curves) separated by $\sim$ 2 hr each (total time span of 8 hr). (Right) Comparison of derived cloud settling rates (shown as altitude of peak density-scaled opacity) to models with different particle sizes.}
    \label{fig:fig14}
\end{figure}

\subsection{Targeted Polar Cloud Observations}

To better isolate individual clouds and track precipitation rates, MCS performed a targeted observation campaign in the north polar region from December 2014 through early 2017. On each pass over the northern winter pole, MCS scanned to observe the same longitude and latitude location, based on the tangent point of the boresight. Several sites of interest were investigated, including the exact pole. CO$_2$ ice opacity profiles were then generated using the CO$_2$ ice/dust opacity scaling factor reported in \citeA{garybicas_asymmetries_2020}. Note that targeted south polar observations were not performed due to operational constraints. Examples of retrieved opacity profiles over the north pole are shown in Figure \ref{fig:fig14}. We first derived the density-scaled opacity based on a constant scale height $H = 10$ km, then found the time-varying altitude of the peak. Comparison of the results to modeled settling rates (Equation \ref{equation:eqwset}) shows agreement for particle radius $r \sim 50~\mu\mathrm{m}$. Thus, the targeted polar cloud observations provide another independent verification of the CO$_2$ snow cloud particle size derived from spectroscopy (Section \ref{subsec:bestfit}).

\section{Conclusions} \label{sec:Conc}

The addition of $\text{CO}_{\text{2}}$ ice opacity profiles to the standard Mars Climate Sounder Level-2 data product is anticipated to provide a comprehensive, multi-year record capable of tracing interannual patterns and variability in $\text{CO}_{\text{2}}$ snowfall within the Martian polar regions. This dataset is of particular importance when considering problems related to the global $\text{CO}_{\text{2}}$ cycle on Mars, polar energy balance, and the stability and growth of polar ice deposits, and has the potential to significantly reshape our understanding of the Martian polar regions as a whole. Here, we present a set of narrow constraints on the physical characteristics of $\text{CO}_{\text{2}}$ ice cloud particles in both the northern and southern polar regions of Mars, which will be essential for future work adding $\text{CO}_{\text{2}}$ ice to the MCS retrieval algorithm.

The primary results of this study are summarized below:

\begin{enumerate}
    \item $\text{CO}_{\text{2}}$ ice clouds over the south pole of Mars are composed of particles that follow a narrow size distribution with $r_{\text{eff}} = 46 \ \mu m$ and $v_{\text{eff}} = 2.0 \times 10^{-3}$, consistent with previous estimates of $\text{CO}_{\text{2}}$ ice cloud particle size on Mars \cite{hayne_carbon_2012, maattanen_troposphere--mesosphere_2022}.
    \item Our results are insensitive to the small water ice cores that are expected to be present within some $\text{CO}_{\text{2}}$ ice cloud particles on Mars \cite{LUGININ2024116271}.
    \item We cannot rule out the possibility that $\text{CO}_{\text{2}}$ ice on Mars may follow a bimodal size distribution with two effective radii peaks at $r_{\text{eff}} = 9 \ \mu m$ and $r_{\text{eff}} = 44 \ \mu m$, though any improvement this has on the goodness of fit of our model is outweighed by the added complexity of dealing with multiple particle size distributions.
    \item The low effective variance of our best-fit size distribution is well-explained by the rapid growth of small ice crystals in the $\text{CO}_{\text{2}}$-rich Martian atmosphere which effectively limits their minimum size, coupled with the rapid sedimentation of large ice crystals out of the atmosphere which limits their maximum size. Together, these processes work to create a strongly peaked particle size distribution.
    \item Particle sizes in the north polar region are very similar to those in the south, with an effective radius of 42 $\mu$m, and an effective variance of $2.0 \times 10^{-3}$. This may be due, in part, to the greater number of condensation nuclei available during the dusty northern winters, which could limit particles from growing much larger, despite the higher surface pressures in the north. It may also be a consequence of the fact that $\text{CO}_{\text{2}}$ ice clouds tend to form near the same pressure level in each hemisphere, making the higher surface pressures in the north irrelevant.
\end{enumerate}

The differences between polar and mesospheric $\text{CO}_{\text{2}}$ ice clouds are worth noting. Mesospheric $\text{CO}_{\text{2}}$ ice clouds are composed of much smaller particles, between $\sim 0.5 - 2.0$ $\mu$m during the day, to upwards of $\sim 7.0$ $\mu$m at night \cite{LUGININ2024116271, CLANCY2019246}. While polar $\text{CO}_{\text{2}}$ ice clouds form in a more stable environment where temperatures naturally fall below the $\text{CO}_{\text{2}}$ frost point for extended periods of time, mesospheric $\text{CO}_{\text{2}}$ clouds form in response to temperature perturbations caused by gravity waves \cite{slipski_mesotides_2022}. Temperatures low enough to support $\text{CO}_{\text{2}}$ nucleation in the equatorial mesosphere are therefore comparatively short-lived, limiting how large $\text{CO}_{\text{2}}$ ice particles can reasonably grow. Interestingly, despite this difference in particle size, \citeA{CLANCY2019246} found evidence of iridescence in some mesospheric $\text{CO}_{\text{2}}$ ice cloud spectra, suggesting a very narrow size distribution $( V_{\text{eff}} \sim 0.03 )$, similar to the ones observed for polar $\text{CO}_{\text{2}}$ ice clouds in this work.

Future work will involve incorporating the infrared scattering parameters for $\text{CO}_{\text{2}}$ ice that are consistent with the physical properties derived in this work (see Table \ref{tab:tab3} in the Appendix) into the MCS retrieval algorithm. This will allow us to begin to generate retrievals of atmospheric $\text{CO}_{\text{2}}$ ice opacity within the polar regions, which is the ultimate goal of this work. The much smaller particle size and more limited lateral extent of mesospheric $\text{CO}_{\text{2}}$ clouds make their inclusion in this new retrieval framework unlikely. We anticipate that this new version of the MCS retrieval algorithm will be capable of retrieving dust, water ice, and $\text{CO}_{\text{2}}$ ice simultaneously when appropriate; however, more work must be done to fully understand the challenges associated with incorporating an entirely new aerosol into this specific retrieval framework. Once completed, this new retrieval version will be applied to the entire MCS dataset to provide a comprehensive climatology of polar $\text{CO}_{\text{2}}$ ice clouds on Mars.

\appendix

\section{Sensitivity Analysis} \label{apsec:Sense}

\begin{table}[b!]
\caption{Best-fit shape parameters including estimated uncertainties in each MCS aerosol channel.} \label{tab:tab2}
\begin{center}
\resizebox{\textwidth}{!}{%
\begin{tabular}{  >{\centering\arraybackslash}p{3cm}  >{\centering\arraybackslash}p{3.3cm}  >{\centering\arraybackslash}p{3.3cm}  >{\centering\arraybackslash}p{3.3cm}  }
\hline
Shape Parameter & Best-Fit Value in A4 & Best-Fit Value in B1 & Best-Fit Value in B2\\ 
\hline
$r_{\text{eff}}$ & $46 \pm 57 \ \mu m$ & $46 \pm 1 \ \mu m$ & $46 \pm 2 \ \mu m$\\
\hline
$v_{\text{eff}}$ & $2.0 \times 10^{-3} \pm 4.1 \times 10^{-3}$ & $2.0 \times 10^{-3} \pm 1.7 \times 10^{-3}$ & $2.0 \times 10^{-3} \pm 2.9 \times 10^{-3}$\\
\hline
\end{tabular}%
}
\end{center}
\end{table}

\begin{wrapfigure}{l}{.62\linewidth}
\includegraphics[scale=.5]{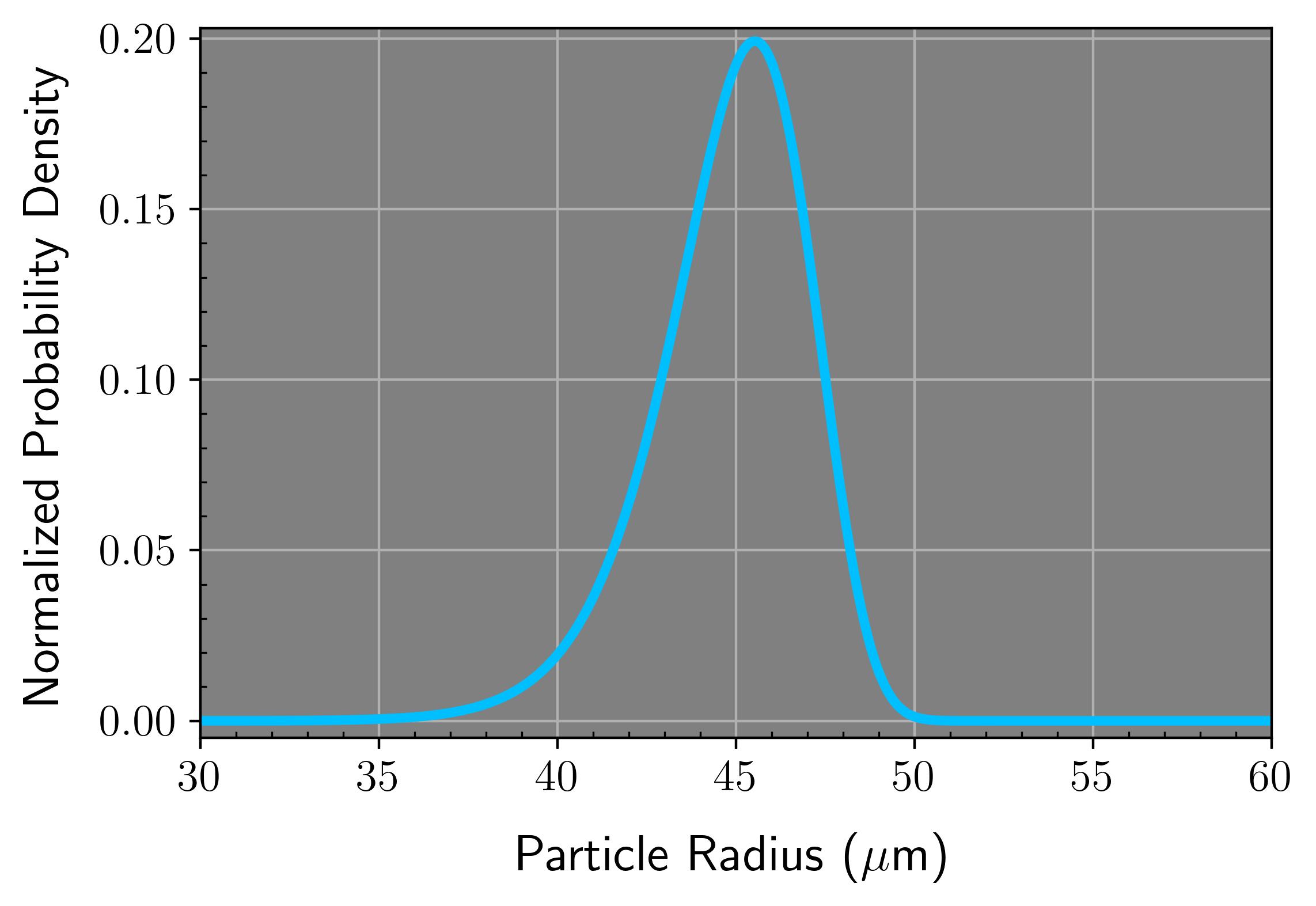}
\caption{The best-fit $\text{CO}_{\text{2}}$ ice size distribution derived in this work.
\label{fig:figA1}}
\end{wrapfigure}

Here, we estimate the confidence in our final results based on an estimation of parameter uncertainties. We do this by conducting a sensitivity analysis of our model to both $r_{\text{eff}}$ and $v_{\text{eff}}$, which describe our best-fit particle size distribution. If we hold one of these parameters constant while allowing the other to change freely, the variance in our set of calculated radiances is given by
\begin{equation} \label{equation:eq22}
    \frac{\sigma_{R}^2}{N} = \sigma_{x}^2 \left( \frac{\partial R \left( x \right)}{\partial x} \right)^2
\end{equation}
\noindent where $N$ is the total number of observations (i.e. the number of calculated radiances in our dataset), $\sigma_R^2$ is the variance in modeled radiances, $R$, $\sigma_x^2$ is the variance in free parameter $x$, and $\frac{\partial R \left( x \right)}{\partial x}$ is the change in modeled radiance with respect to $x$. Solving for $\sigma_x$, we obtain:
\begin{equation} \label{equation:eq23}
    \sigma_{x} = \frac{\sigma_{R}}{\sqrt{N}} \left( \frac{\partial R \left( x \right)}{\partial x} \right)^{-1}
\end{equation}
Because we have a different set of modeled radiances for each aerosol channel, our final uncertainties are also channel-dependent. Best-fit shape parameters and associated uncertainties are provided in Table \ref{tab:tab2}, corresponding best-fit Mie parameters for each MCS channel are given in Table \ref{tab:tab3}, and the best-fit normalized size distribution is plotted in Figure \ref{fig:figA1}. 
It should be noted that the error in channel A4 is quite large, which is consistent with the fact that A4 has the lowest signal-to-noise out of the three aerosol channels we considered. Unfortunately, this means that A4 provides little leverage for constraining $r_{\text{eff}}$ and $v_{\text{eff}}$, to the extent that our best-fit values are only upper limits in that channel. Luckily, B1 and B2 are more sensitive to changes in $r_{\text{eff}}$ and $v_{\text{eff}}$ such that our constraint on particle size, especially, is quite narrow. Our best-fit effective variance is much more loosely constrained in all three channels, and should realistically be taken as an upper limit. That being said, we find it unlikely that $\text{CO}_{\text{2}}$ ice clouds form even narrower size distributions given that, as mentioned above, expanding our parameter space to include even smaller values of $v_{\text{eff}}$ does little to improve our model fit.

\begin{figure}[b!]
\centering
\hspace{-.7cm}\includegraphics[width=.93\textwidth]{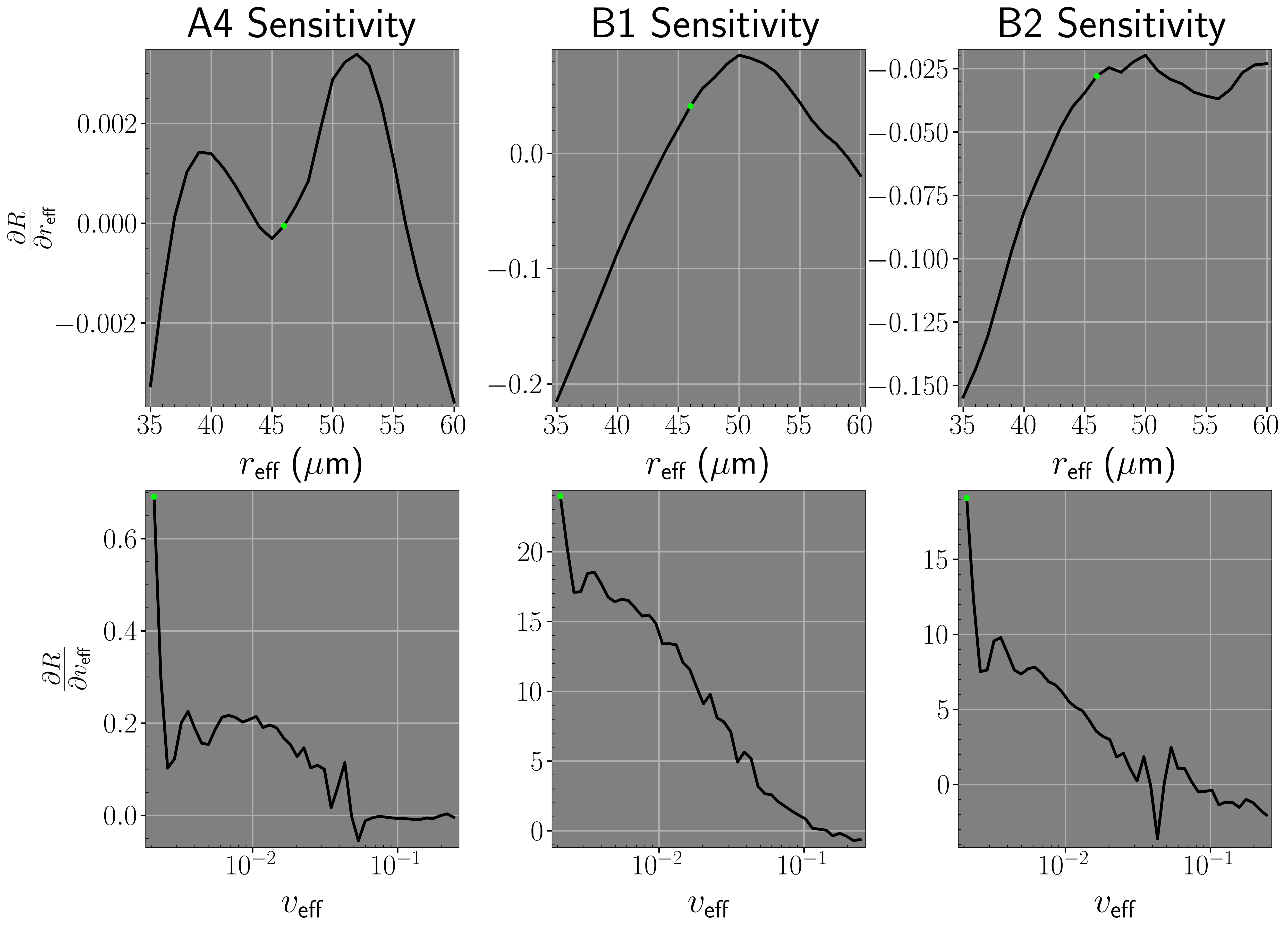}
\caption{Results of the sensitivity analysis showing the change in modeled radiance with respect to each shape parameter while holding the other shape parameters constant. Plots in each row show sensitivity to a different shape parameter, while columns represent results from the three different MCS aerosol channels considered in this work. Green dots indicate the best-fit values reported in Table \ref{tab:tab2}.
\label{fig:figA2}}
\end{figure}

Our calculation of $\frac{\partial g \left( x \right)}{\partial x}$ from Equation \ref{equation:eq23} also allows us to better understand how our final results depend on changes in each of our model’s free parameters. Figure \ref{fig:figA2} shows that calculated radiances are most sensitive to $r_{\text{eff}}$ near our best-fit value; any small change in $r_{\text{eff}}$ will therefore result in a significant change in the modeled radiances, indicating our constraint on $r_{\text{eff}}$ is particularly robust.

\pagebreak

\begin{table}[t!]
\caption{Derived Mie parameters using the best-fit size distribution given in Section \ref{subsec:bestfit}. Center frequencies for the B channels are detector dependent--here they are provided for detector 11 which falls in the middle of the sensor array.} \label{tab:tab3}
\begin{center}
\resizebox{\textwidth}{!}{%
\begin{tabular}{  >{\centering\arraybackslash}p{1.5cm}  >{\centering\arraybackslash}p{5cm}  >{\centering\arraybackslash}p{1.5cm}  >{\centering\arraybackslash}p{1.5cm}  >{\centering\arraybackslash}p{1.5cm}  }
\hline
Channel & Center Frequency $\left( \text{cm}^{-1} \right)$ & $Q_{ext}$ & $\varpi_0$ & $g$ \\ 
\hline
A1 & 606.916 & 2.378 & 0.8906 & 0.7833 \\
\hline
A2 & 631.017 & 2.249 & 0.7913 & 0.7971 \\
\hline
A3 & 648.703 & 2.271 & 0.6172 & 0.8523 \\
\hline
A4 & 842.724 & 2.254 & 0.9882 & 0.8061 \\
\hline
A5 & 463.436 & 2.569 & 0.9993 & 0.7682 \\
\hline
B1 & 315.924 & 1.980 & 0.9986 & 0.5804 \\
\hline
B2 & 253.553 & 2.741 & 0.9958 & 0.6641 \\
\hline
B3 & 242.144 & 2.774 & 0.9960 & 0.6743 \\
\hline
\end{tabular}%
}
\end{center}
\end{table}

\pagebreak

\section*{Data Availability Statement}

All data used in this analysis, including the initial set of MCS $\text{CO}_{\text{2}}$ ice retrievals, are publicly available in an online repository \cite{stevens_2025_data}. The code used to generate the results and figures in this work, as well as the simplified convective cooling model, are also publicly archived online \cite{stevens_2025_code}.

\acknowledgments

The authors would like to thank Mike Wolff as well as the Mars Climate Sounder team for helpful discussions that improved the quality of this manuscript. This work was supported by the National Aeronautics and Space Administration through the Mars Reconnaissance Orbiter project. Work at the Jet Propulsion Laboratory, California Institute of Technology, is performed under contract with the National Aeronautics and Space Administration (80NM0018D0004).

%
%

\bibliography{Bib}

%
%
%
%
%

\end{document}